\documentclass[12pt,a4paper]{article}

\usepackage[T1]{fontenc}
\usepackage[latin9]{inputenc}
\usepackage[a4paper]{geometry}
\usepackage{amsmath}
\usepackage{amsfonts}
\usepackage{amssymb} 
\usepackage{graphicx}
\usepackage{a4wide}
\usepackage{hepunits}
\usepackage{url}
\usepackage{supertabular}
\usepackage{xspace}
\usepackage{latexsym}
\usepackage{textcomp}
\usepackage{wrapfig}
\usepackage{subfig}
\usepackage{slashed}
\usepackage{hepunits}
\usepackage{setspace}

\DeclareMathOperator{\sgn}{sgn}
\DeclareMathOperator{\SymCovDev}{\overset{\leftrightarrow}D}
\DeclareMathOperator{\RightCovDev}{\overset{\rightarrow} D}
\DeclareMathOperator{\LeftCovDev}{\overset{\leftarrow}D}

\newcommand{\refeq}[1]{Eq.~(\ref{#1})}
\newcommand{\reftab}[1]{Tab.~\ref{#1}}
\newcommand{\reffig}[1]{Fig.~\ref{#1}}
\newcommand{\refcite}[1]{Ref.~\cite{#1}}
\newcommand{\refsec}[1]{Sec.~\ref{#1}}

\newcommand{\ie}{\textit{i.e. }}

\newcommand{\etal}{\textit{et al. }}

\newcommand{\xb}{x_B}

\newcommand{\ImH}{Im \mathcal{H}}
\newcommand{\ReH}{Re \mathcal{H}}

\newcommand{\ket}[1]{\ensuremath{\left|#1\right\rangle}\xspace}
\newcommand{\bra}[1]{\ensuremath{\left\langle #1\right|}\xspace}

\begin{document}
\begin{flushright}
IRFU-13-94

April, 29 2013\\[10mm]
\end{flushright}


\begin{center}
\begin{spacing}{1.5}

{\Large\bf Test of two new parameterizations \\ of the Generalized Parton Distribution $H$} \\
\end{spacing}
\vskip 10mm

C.~M\scshape {ezrag}\footnote{cedric.mezrag@cea.fr}, H.~M\scshape {outarde}\footnote{herve.moutarde@cea.fr} , F.~S\scshape {abati\'e}\footnote{franck.sabatie@cea.fr}
\\[1em]
{\small {\it IRFU/Service de Physique Nucl\'eaire \\ CEA Saclay, F-91191 Gif-sur-Yvette, France}}\\

\end{center}
\vskip 8mm 
\begin{abstract}
In 2011 Radyushkin outlined the practical implementation of the One-Component Double Distribution formalism in realistic Generalized Parton Distribution models. We compare the One-Component Double Distribution framework to the standard one and compute Deeply Virtual Compton Scattering observables for both. In particular the new implementation is more flexible, offering a greater range of variation of the real and imaginary parts of the associated Compton Form Factor while still allowing to recover results similar to the classical approach. Moreover the polynomiality property is satisfied up to the highest order. Although the comparison to experimental data may be improved, the One-Component Double Distribution modeling is thus an attractive alternative. 
\end{abstract}

\vskip 10mm


\section*{Introduction}

Generalized Parton Distributions (GPDs) were introduced by M\"uller \etal \cite{Mueller:1998fv}, Ji \cite{Ji:1996nm} and Radyushkin \cite{Radyushkin:1997ki}. They encode a wealth of information about the structure of the nucleon including 3D imaging of its partonic content and access to the quark orbital angular momentum. GPDs have been the object of an intense theoretical and experimental activity ever since (see the reviews \refcite{Ji:1998pc, Goeke:2001tz, Diehl:2003ny, Belitsky:2005qn, Boffi:2007yc,Guidal:2013rya} and references therein). 

GPDs can be accessed experimentally by studying the processes of leptoproduction of a photon: Deeply Virtual Compton Scattering (DVCS) (GPDs can also be accessed through its timelike counterpart: Timelike Compton Scattering \cite{Berger:2002}), or a meson: Deeply Virtual Meson Production (DVMP). The DVCS amplitude is expressed in terms of Compton Form Factors (CFFs), which are convolutions of GPDs with known kernels. CFFs were extracted from DVCS in \refcite{Guidal:2008ie, Guidal:2009aa, Guidal:2010de, Guidal:2010ig, Kumericki:2011rz, Kumericki:2013br, Kumericki:2009uq, Moutarde:2009fg} while GPDs were constrained from DVCS in \refcite{Goldstein:2010gu} or from DVMP in \refcite{Goloskokov:2005sd, Goloskokov:2007nt, Goloskokov:2009ia}. Note that this last set of GPDs, tuned for DVMP analysis, has been compared recently to almost all existing DVCS measurements \cite{Kroll:2012sm}. Although these results are promising, our knowledge of GPDs is far from complete and will certainly be challenged by forthcoming measurements at COMPASS, by the high precision data expected after the Jefferson Lab 12~\GeV\  upgrade and at an Electron Ion Collider. Gaining an experimental knowledge of GPDs is more involved than extracting Parton Distribution Functions (PDFs) from measurements: there are more GPDs than PDFs, and they depend on three variables instead of one. GPDs are subject to several theoretical constraints, and are related to PDFs and quark contributions to nucleon Form Factors (FFs). There is no known parameterization of GPDs relying only on QCD first principles. 

Part of the theoretical constraints on GPDs are fulfilled by modeling Double Distributions (DDs) \cite{Mueller:1998fv, Radyushkin:1998es, Radyushkin:1998bz}, which are related to GPDs by a Radon transform \cite{Teryaev:2001qm}. DD modeling is the most popular way to build simple and realistic GPD models, and has been widely used so far, for example in the popular VGG (Vanderhaeghen, Guichon and Guidal) \cite{Goeke:2001tz, Vanderhaeghen:1998uc, Guichon:1998xv, Vanderhaeghen:1999xj, Guidal:2004nd} or GK (Goloskokov and Kroll) \cite{Goloskokov:2005sd, Goloskokov:2007nt, Goloskokov:2009ia} models. However these models are not in agreement with all existing DVCS data, as one can see from fit results for VGG \cite{Guidal:2008ie, Guidal:2009aa, Guidal:2010de, Guidal:2010ig} or in the systematic application of the GK model to DVCS data in \refcite{Kroll:2012sm}. To reach a better comparison to the data one may either turn to more sophisticated GPD parameterizations, for example as advocated in \refcite{Kumericki:2008di}, or try different implementations of DD modeling. The latter direction recently benefited from renewed interest when Radyushkin exhibited a GPD model \cite{Radyushkin:2011dh} built on the so-called \emph{One-Component DDs} representation \cite{Belitsky:2005qn}. This representation was first outlined in \refcite{Belitsky:2000vk} but was not used for phenomenology so far: its implementation was an open problem since it leads to a more divergent behavior of the GPDs at small longitudinal momentum fractions than the one described in \refcite{Radyushkin:1998es, Radyushkin:1998bz}.

In this paper we implement the One-Component DD formalism along the lines described in \refcite{Radyushkin:2011dh} for PDFs with an integrable singularity at the origin, and confront it to DVCS experimental data. We will compare our model to Jefferson Lab Hall~A \cite{Munoz Camacho:2006hx} and CLAS measurements \cite{Girod:2007aa} which are very accurate and concern the valence sector. Therefore we only implemented the One-Component DD model for the singlet contribution of valence quarks. Since the partonic interpretation of DVCS observables relies on factorization theorems, we will apply the further restriction $| t | / Q^2 \lesssim 0.1$ where $t$ is the momentum transfer, and $Q^2$ the incoming photon virtuality \cite{Radyushkin:1997ki, Radyushkin:1996nd, Ji:1998xh, Collins:1998be}. The work of \refcite{Radyushkin:2011dh} was done in the case of a spinless target although it was illustrated by a nucleon valence PDF-like toy model. The analog of the GPD $H_{\textrm{spin} 0}$ in the spinless case is an admixture of the GPDs $H_{\textrm{spin} 1/2}$ and $E_{\textrm{spin} 1/2}$. As discussed in \refcite{Kroll:2012sm}, beam polarized DVCS observables involving unpolarized targets are mostly independent of the GPD $E$. This allows us to use the results of \refcite{Radyushkin:2011dh} without any further adaptation to the spin-1/2 case in this first study. 

In the first section we remind the basics of GPD modeling based on DDs. In the second section we modify accordingly the valence part of the DD model used in \refcite{Kroll:2012sm} and in the third section we discuss the phenomenological consequences of this alternative DD Ansatz. In the fourth section we discuss the model for nucleon GPDs relying on both 1CDD and 2CDD formalisms recently advocated in \refcite{Radyushkin:2013hca}.


\section{GPD models in the Factorized Double Distribution approach}

For any four-vector $v$ we note:
\begin{equation}
\label{eq-def-lc-coordinates}
v^{\pm} = \frac{1}{\sqrt{2}} ( v^0 \pm v^3 ) \quad \textrm{ and } \quad v = ( v^+, v_\perp, v^-).
\end{equation}
$(uv)$ denotes the scalar product of two four-vectors $u$ and $v$.

\subsection{One-Component DD and Two-Component DD formalisms}
\label{sec:1CDD-2CDD-spinless-target}

For the sake of clarity, we discuss in this section the case of GPDs and DDs of spinless hadrons. 

The GPD $H^q$, $q$ denoting the quark flavor, is introduced through the matrix element of \refeq{eq-def-GPD-H-spinless-target}:
\begin{equation}
\label{eq-def-GPD-H-spinless-target}
H^q( x, \xi, t ) =  \frac{1}{2} \int \frac{\mathrm{d}z^-}{2\pi} \, e^{i x P^+ z^-} \bra{P+\frac{\Delta}{2}} \bar{q}\left( -\frac{z}{2} \right) \gamma^+ q \left( \frac{z}{2} \right) \ket{P-\frac{\Delta}{2}}_{z^+=0 \atop z_\perp=0}, 
\end{equation}
where  $\xi = - \Delta^+/(2P^+)$ is the skewness.

The DDs $F^q$ and $G^q$ of the two-component DD (2CDD) formalism associated to the quark flavor $q$ are defined by the following matrix element:
\begin{eqnarray}
\bra{P+\frac{\Delta}{2}} \bar{q}\left( -\frac{z}{2} \right) \gamma_\mu q \left( \frac{z}{2} \right) \ket{P-\frac{\Delta}{2}}_{z^2=0} 
& = & 2P_{\mu}\int_{\Omega} \mathrm{d}\beta\mathrm{d}\alpha \, e^{- i \beta (P z) + i \alpha \frac{(\Delta z)}{2}} F^q( \beta, \alpha, t ) \nonumber \\
& & \, - \Delta_{\mu}\int_{\Omega} \mathrm{d}\beta\mathrm{d}\alpha \, e^{- i \beta (P z) + i \alpha \frac{(\Delta z)}{2}} G^q( \beta, \alpha, t ) \nonumber \\
& & +\text{ higher twist terms},
\label{eq-def-DD-F-G-spinless-target}
\end{eqnarray}
where $\bar{q}$ and $q$ denote the quark fields separated by the light-like distance $z$, $P \pm \Delta/2$ the momenta of the incoming (+) and outgoing (-) hadrons and $t = \Delta^2$ is the usual Mandelstam variable. When not necessary, the $t$-dependence will not be explicitly written. $\Omega$ is the rhombus defined by: 
\begin{equation}
\label{eq-def-rhombus-DD}
|\alpha| + |\beta| \leq 1. 
\end{equation}
This yields the following relation between GPDs and DDs:
\begin{equation}
H^q( x, \xi, t ) = \int_\Omega \mathrm{d}\beta\mathrm{d}\alpha \, \delta( x - \beta - \alpha \xi ) \big( F^q( \beta, \alpha ) + \xi G^q( \beta, \alpha, t ) \big).
\label{eq-relation-DD-GPD}
\end{equation}

Assuming the vanishing of DDs on the boundary of the rhombus\footnote{As stated in \refcite{Tiburzi:2004qr} DDs need only vanish at the corners $(\beta = \pm 1, \alpha = 0)$ and $(\beta = 0, \alpha = \pm 1)$ of the support $\Omega$. This fact introduces an extra boundary condition which does not modify our discussion and thus is omitted.} $\Omega$, one can see that \emph{gauge transforming}\footnote{The terminology of \emph{gauge tranformation} has been chosen in \refcite{Teryaev:2001qm} for its formal likeness with the gauge transformation of the vector potential of the static two-dimensional magnetic field.}  \cite{Teryaev:2001qm, Tiburzi:2004qr} the DDs $F$ and $G$ by means of an arbitrary function $\sigma^q$:
\begin{eqnarray}
F^q(\beta,\alpha) & \rightarrow & F^q(\beta,\alpha) + \frac{\partial \sigma^q}{\partial \alpha}(\beta,\alpha),  \label{eq-def-gauge-transform-F} \\
G^q(\beta, \alpha) & \rightarrow & G^q(\beta, \alpha) - \frac{\partial \sigma^q}{\partial \beta }(\beta,\alpha),  \label{eq-def-gauge-transform-G} 
\end{eqnarray}
 does not change the matrix element on the left-hand side of \refeq{eq-def-DD-F-G-spinless-target}, neither does it modify the resulting GPD in \refeq{eq-relation-DD-GPD}. From time reversal invariance $G^q$ and $\sigma^q$ are $\alpha$-odd while $F^q$ is $\alpha$-even. 

A representation relying on one unique DD $f^q( \beta, \alpha )$ was proposed in \refcite{Belitsky:2000vk} and is named One-Component DD (1CDD) in \refcite{Belitsky:2005qn}. The non-trivial $\beta$ and $\alpha$ dependence of the $F$ and $G$ type DDs is expressed in terms of $f$ by the following:
\begin{eqnarray}
F^q_{1\textrm{CDD}}( \beta, \alpha ) & = & \beta f^q( \beta, \alpha ), \label{eq-F-1CDD} \\
G^q_{1\textrm{CDD}}( \beta, \alpha ) & = & \alpha f^q( \beta, \alpha ). \label{eq-G-1CDD}
\end{eqnarray}
This DD representation is less known than the 2CDD but is equivalent as it can be obtained from the 2CDD representation by a gauge transform (see \refcite{Tiburzi:2004qr}). In the 1CDD representation \refeq{eq-def-DD-F-G-spinless-target} can thus be written:
\begin{eqnarray}
\bra{P+\frac{\Delta}{2}} \bar{q}\left( -\frac{z}{2} \right) \gamma_\mu q \left( \frac{z}{2} \right) \ket{P-\frac{\Delta}{2}}_{z^2=0} 
& = & \int_{\Omega} \mathrm{d}\beta\mathrm{d}\alpha \, e^{- i \beta (P z) + i \alpha \frac{(\Delta z)}{2}} \Big( 2P_{\mu} \beta - \Delta_\mu \alpha \Big) f^q( \beta, \alpha, t ) \nonumber \\
& & +\text{ higher twist terms}.
\label{eq-def-DD-f-spinless-target-1CDD}
\end{eqnarray}

With our forthcoming application to DVCS data in mind, let us focus on singlet GPDs and DDs, defined by:
\begin{eqnarray}
H^{q+}( x, \xi ) & = & H^q( x, \xi ) - H^q( -x, \xi ), \label{eq:def-singlet-gpd-H} \\
F^{q+}( \beta, \alpha ) & = & F^q( \beta, \alpha ) - F^q( -\beta,  \alpha ), \label{eq:def-singlet-dd-F} \\
G^{q+}( \beta, \alpha ) & = & G^q( \beta, \alpha ) + G^q( -\beta,  \alpha ). \label{eq:def-singlet-dd-G}
\end{eqnarray}
The $\beta$-odd $\sigma_D$ function:
\begin{equation}
\label{eq-sigma-D-two-DD-to-DD-plus-D}
\sigma^q_D(\beta,\alpha) = -\frac{1}{2} \left[ \int_{-1+|\alpha|}^{\beta} d\mathrm{\beta'} G^{q+}( \beta', \alpha ) - \int_{\beta}^{1 - |\alpha|} d\mathrm{\beta'} G^{q+}( \beta', \alpha ) - \sgn( \beta ) D^q( \alpha ) \right],
\end{equation}
reduces the $G$-type singlet DD $G_{\textrm{DD+D}}^+( \beta, \alpha )$ to a function $\delta( \beta ) D( \alpha )$ with a trivial dependence on the variable $\beta$: the so-called Polyakov~-~Weiss $D$-term \cite{Polyakov:1999gs} while all the $\beta$ dependence goes to the $F$-type singlet DD $F_{\textrm{DD+D}}^+( \beta, \alpha )$. This representation is coined ``DD+D'' in \refcite{Radyushkin:2011dh} and corresponds to the \emph{Polyakov~-~Weiss gauge}.

A popular way to model the DD $F^q$ is Radyushkin's \emph{Factorized Double Distribution Ansatz} (FDDA) \cite{Musatov:1999xp} which makes contact with the forward limit $q( x )$ of the GPD $H^q( x )$ when $t$ = 0, or with the (Fourier transformed) unintegrated forward limit $q( x, t )$ when $t \neq 0$:
\begin{equation}
F^{q}( \beta, \alpha, t ) = \pi_N( \beta, \alpha ) q( \beta, t ),
\label{eq-fdda}
\end{equation}
where the profile function $\pi_N$ reads:
\begin{equation}
\label{eq-def-profile-function}
\pi_N( \beta, \alpha ) = \frac{\Gamma\left( N + \frac{3}{2} \right)}{\sqrt{\pi} \Gamma( N + 1)} \frac{[ ( 1 - |\beta| )^2 - \alpha^2 ]^N}{( 1 - |\beta| )^{2 N + 1}}.
\end{equation}
$\Gamma$ denotes the Euler Gamma function. The coefficient $\Gamma( N + 3/2) / (\sqrt{\pi} \Gamma( N + 1))$ guarantees the normalization of the profile function:
\begin{equation}
\label{eq-profile-function-normalization}
\int_{-1 + |\beta|}^{+1 - |\beta|} \mathrm{d}\alpha \, \pi_N( \beta, \alpha ) = 1~.
\end{equation}
From a matter of principles, the FDDA can be applied to a $F$-type DD in the 1CDD and 2CDD formalisms as well. However the FDDA breaks the equivalence between the two representations as we can see by implementing the gauge transformation between the DD+D and the 1CDD representation: from \refeq{eq-def-gauge-transform-F} we see that:
\begin{eqnarray}
\partial \sigma / \partial \alpha 
& = & 
F^q_{\textrm{1CDD}}(\beta, \alpha) -  F^q_{\textrm{DD+D}}(\beta,\alpha) \nonumber \\
& = & 
\pi_N(\beta,\alpha)q(\beta) - \pi_N(\beta,\alpha)q(\beta)  \nonumber \\
& = & 
0.  \label{eq:vanishing-sigma}
\end{eqnarray}
and $\sigma$ does not depend on $\alpha$; Being constant and $\alpha$-odd, $\sigma$ must be zero. But $\sigma_D$ in \refeq{eq-sigma-D-two-DD-to-DD-plus-D} is manifestly non-trivial. Choosing to apply the FDDA to one representation or another is thus a phenomenological choice and can \textit{a priori} be decided by comparison to experimental data.

One of the advantages of the 1CDD parameterization involves the polynomiality of Mellin moments:
\begin{equation}
\int_{-1}^{1} \mathrm{d}x\ x^nH^q(x,\xi) = \sum_{i=0 \atop {\rm even}}^{n}(2\xi )^i A_{n+1,i}^q + \bmod(n,2)(2\xi )^{n+1}C^q_{n+1}, \label{eq:Mellin}
\end{equation}
where the scalar coefficients $A_{n+1,i}^q$ and $C^q_{n+1}$ depend on $t$. Indeed, in the 2CDD (DD+D) representation, the term of highest degree is generated by the $G$-type DD (D-term), \ie the $F$-type DD alone does not fulfill the polynomiality property. On the contrary, in the 1CDD representation, all powers of $\xi$ are correctly generated by one DD only.

Note however that the FDDA applied to the 1CDD formalism leads to more singular GPDs. Indeed, in that case the factorized Ansatz yields:
\begin{equation}
f^{q}( \beta, \alpha ) = \frac{q( \beta )}{\beta} \pi_N( \beta, \alpha ). 
\label{eq-FDDA-1CDD}
\end{equation}
The relation in \refeq{eq-relation-DD-GPD} between the GPD $H^{q}$ and the DDs in the 1CDD framework (see \refeq{eq-F-1CDD}, \refeq{eq-G-1CDD} and \refeq{eq-FDDA-1CDD}) reads:
\begin{eqnarray}
H^{q}( x, \xi ) 
& = & \int_\Omega \mathrm{d}\beta\mathrm{d}\alpha \, \delta( x - \beta - \alpha \xi ) \big(  F^{q}_{\textrm{1CDD}}( \beta, \alpha ) + \xi G^{q}_{\textrm{1CDD}}( \beta, \alpha ) \big) \nonumber \\
& = & \int_\Omega \mathrm{d}\beta\mathrm{d}\alpha \, \delta( x - \beta - \alpha \xi ) (  \beta  + \alpha \xi  ) f^{q}( \beta, \alpha ) \nonumber \\
& = & 
x \int_\Omega \mathrm{d}\beta\mathrm{d}\alpha \, \delta( x - \beta - \alpha \xi ) \frac{q( \beta )}{\beta} \pi_N( \beta, \alpha ).
\label{eq-GPD-H-1CDD-FDDA}
\end{eqnarray}
Given the typical behavior of nucleon valence PDFs $q_{\textrm{val}}( \beta ) \propto \beta^{-0.5}$ for small $\beta$, the divergence at $\beta = 0$ in \refeq{eq-GPD-H-1CDD-FDDA} is problematic. Looking at sea quarks, the divergence becomes even worse, as the behavior of the associated PDFs is given by $q_{\textrm{sea}}(\beta) \propto \beta^{-(1+\delta)}$ with $0 < \delta < 1$, when $\beta$ goes to $0$.

To unambiguously define the valence and sea contributions to PDFs and GPDs, we follow the conventions of \refcite{Goloskokov:2006hr} and references therein:
\begin{itemize}
\item The valence and sea contributions $q_{\textrm{val}}$ and $q_{\textrm{sea}}$ to the PDF $q$ are defined on $[-1, +1]$ by:
\begin{eqnarray}
q_{\textrm{val}}( \beta ) & = & \theta( \beta ) q_{\textrm{val}\,|[0, 1]}( \beta ), \label{eq:def-qval-negative-beta} \\
q_{\textrm{sea}}( \beta ) & = & \sgn( \beta ) q_{\textrm{sea}\,|[0, 1]}( | \beta | ), \label{eq:def-qsea-negative-beta} 
\end{eqnarray}
where $q_{|[0, 1]}$ denotes the restriction of the PDF $q$ to the interval $[0, 1]$.
\item The valence and sea contributions $F^q_{\textrm{val}}$ and $F^q_{\textrm{sea}}$ to the DD $F^q$ are:
\begin{eqnarray}
F^q_{\textrm{val}}( \beta, \alpha )
& = & \Big( F^q( \beta, \alpha ) + F^q( - \beta, \alpha ) \Big) \theta( \beta ) \label{eq:def-valence-DD-F} \\
F^q_{\textrm{sea}}( \beta, \alpha )
& = & \Big( F^q( \beta, \alpha ) \theta( \beta ) - F^q( - \beta, \alpha ) \Big) \theta( - \beta ) \label{eq:def-sea-DD-F}
\end{eqnarray}
\end{itemize}
To take care of this decomposition the classical FDDA (\ref{eq-fdda}) writes:
\begin{eqnarray}
F^q_{\textrm{val}}( \beta, \alpha )
& = & \pi_{N_{\textrm{val}}}( \beta, \alpha ) q_{\textrm{val}}( \beta ), \label{eq:FDDA-valence-part} \\
F^q_{\textrm{sea}}( \beta, \alpha )
& = & \pi_{N_{\textrm{sea}}}( \beta, \alpha ) q_{\textrm{sea}}( \beta ), \label{eq:FDDA-sea-part} 
\end{eqnarray}
where the profile function parameters $N_{\textrm{val}}$ and $N_{\textrm{sea}}$ can be distinct numbers.

\subsection{Taming the divergent structure of the One-Component DD formalism}

In this section we describe Radyushkin's solution \cite{Radyushkin:2011dh, Radyushkin:2012gba} to control the potential divergence of \refeq{eq-GPD-H-1CDD-FDDA} and prepare the ground for an implementation into a realistic GPD model. More precisely, Radyushkin made manifest that the 1CDD structure of the GPD model of Szczepaniak \etal derived in \refcite{Szczepaniak:2007af}, and explained how the aforementioned divergent behavior at small parton longitudinal momentum fraction can be controlled through a dispersion relation.

\begin{figure}[h]
	\begin{center}
           \includegraphics[width=0.5\textwidth]{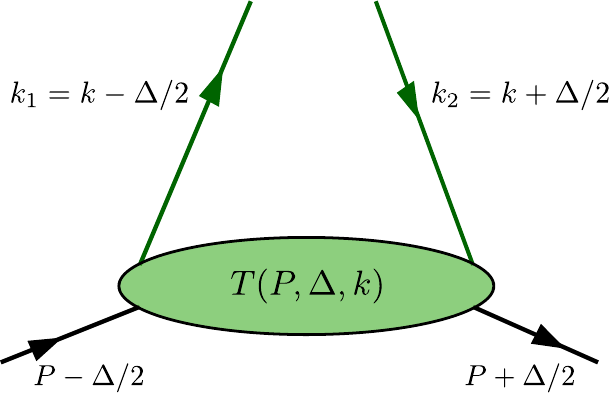}
           \caption{Quark hadron scattering amplitude.}
           \label{fig-quark-hadron-scattering-amplitude}
	\end{center}
\end{figure}

The considered model is essentially the computation of a triangle diagram. It relies on three main assumptions  (see \reffig{fig-quark-hadron-scattering-amplitude} for notations):
\begin{description}
\item[(i)\hphantom{ii}] A parton~-~hadron scattering amplitude $T( P, \Delta, k )$ in replacement of the spectator quark propagator.
\item[(ii)\hphantom{i}] A once-subtracted dispersion relation for the quark-hadron scattering amplitude $T( P, \Delta, k)$:
\begin{equation}
\label{eq-once-subtracted-dispersion-relation}
T( P, \Delta, k ) = T_0( t ) + \int_0^{+\infty} \mathrm{d}\sigma \, \rho( \sigma ) \left[ \frac{1}{\sigma - ( P - k )^2} - \frac{1}{\sigma} \right],
\end{equation}
where the $t$-dependent subtraction constant $T_0$ is unknown and $\rho$ is a spectral function that generates a Regge-behavior for PDFs.
\item[(iii)] A modification of the quark propagators:
\begin{equation}
\frac{1}{( m^2 - k_i^2 )} \rightarrow \frac{1}{( m^2 - k_i^2 )^N} \quad \textrm{for } i = 1, 2.
\end{equation}
\end{description}

The spectral function $\rho$ can be traded for the forward limit $q_{\textrm{val}}( x )$ of the GPD $H^q_{\textrm{val}}( x, \xi )$\footnote{See Eq.~(16) of \refcite{Radyushkin:2012gba}; there is a typing mistake in Eq.~(36) of \refcite{Radyushkin:2011dh} modifying the expression of the regularizing term $\delta[ x - \alpha/(1-\beta) \xi]$.}:
\begin{equation}
\frac{H^q_{\textrm{val}}( x, \xi )}{x} = \int_0^1 \mathrm{d}\beta \, \int_{-1+\beta}^{+1-\beta} \mathrm{d}\alpha \, \frac{q_{\textrm{val}}( \beta )}{\beta} \pi_N( \beta, \alpha ) \left[ \delta( x - \beta - \alpha \xi )- \frac{1}{( 1 - \beta )^2} \delta\left( x - \frac{\alpha \xi}{1-\beta} \right) \right]. 
\label{eq-analogous-eq-48-radyushkin-paper}
\end{equation}
Having remarked that $\pi_N\Big( \beta, \alpha ( 1 - \beta ) \Big) = \pi_N( 0, \alpha ) / ( 1 - \beta )$, a change of variables yields:
\begin{equation} 
\frac{H^q_{\textrm{val}}( x, \xi )}{x}  = \int_0^1 \mathrm{d}\beta \, \int_{-1+\beta}^{+1-\beta} \mathrm{d}\alpha \, \delta( x - \beta - \alpha \xi )  \left[ \frac{q_{\textrm{val}}( \beta )}{\beta} \pi_N( \beta, \alpha ) - \delta( \beta ) \pi_N( 0, \alpha ) \int_0^1 \mathrm{d}\gamma \frac{q_{\textrm{val}}( \gamma )}{\gamma ( 1 - \gamma )^2} \right] \label{eq-analogous-eq-20-radyushkin-proceedings}
\end{equation}
This last equation clearly displays the 1CDD structure of this model, with:
\begin{equation}
f_{\textrm{val}}^q( \beta, \alpha ) = \frac{q_{\textrm{val}}( \beta )}{\beta} \pi_N( \beta, \alpha ) - \delta( \beta ) \pi_N( 0, \alpha ) \int_0^1 \mathrm{d}\gamma \frac{q_{\textrm{val}}( \gamma )}{\gamma ( 1 - \gamma )^2}.
\label{eq:1CDD-structure-szczepaniak-et-al-model}
\end{equation}
The singular behavior of the forward function $q_{\textrm{val}}( \beta ) / \beta$ is regularized by the second term in the brackets of \refeq{eq-analogous-eq-48-radyushkin-paper} as will be shown explicitly below in \refeq{eq-H-1CDD-corrected-mistake-x-small} and \refeq{eq-H-1CDD-corrected-mistake-x-large} when implementing the model. This term stems from the $1/\sigma$ term of the dispersion relation \refeq{eq-once-subtracted-dispersion-relation}. 

At last, the subtraction constant $T_0$ generates an additional term $D_0^q$:
\begin{equation}
\label{eq-additional-DTerm-D0}
D_0^q\left( \frac{x}{\xi}, t \right) = \frac{T^q_0( t )}{2N 2^{2N}} \frac{x}{\xi}  \left( 1 - \frac{x^2}{\xi^2} \right)^N \theta( | x | < \xi ),
\end{equation}
which can be recast in the 1CDD framework by introducing:
\begin{equation}
\label{eq-additional-DTerm-D0-1CDD-picture}
f_0^q( \beta, \alpha ) = \int_{\Omega} \mathrm{d}\beta \mathrm{d}\alpha \, \delta( x - \beta - \alpha \xi ) \delta( \beta ) \frac{D_0^q( \alpha )}{\alpha},
\end{equation}
where the variable $t$ has been omitted for clarity.


\section{Implementation}

As stated in the introduction, the VGG and GK models are probably the DD models that are most often used for phenomenological applications. Both are expressed in the 2CDD formalism, in the specific DD+D representation. Moreover the $D$-term is set to 0 in the GK implementation.

The GK model was built in order to interpret the experimental results on Deeply Virtual Meson Production (DVMP), and as such, it was designed to work at small to intermediate $\xb$ values. It was recently confronted to DVCS measurements in \refcite{Kroll:2012sm}. Without any tuning of parameters, the model reaches a good agreement to the data at small $x_B$ \ie in the sea quark region. Not surprisingly the comparison is less satisfactory in the valence region, leaving open the question of the extension of the model's validity range. A natural question is: Can the agreement be improved in the valence region by switching from the 2CDD to the 1CDD formalism?

\subsection{Valence part of the GPD $H$ in the Goloskokov~-~Kroll model}
\label{sec:H-valence-GK}

The valence GPD $H_{\textrm{val}}^q$ for the quark flavor $q$ is described in the 2CDD formalism, in the DD+D representation, assuming a vanishing D-term. In the following this specific representation will be simply referred to as "DD".  A factorized Ansatz is used:
\begin{equation}
\label{eq-GK-Hval-parameterization}
H_{\textrm{val}}^q( x, \xi, t, \mu^2 ) = \int_\Omega \mathrm{d}\beta\mathrm{d}\alpha \, \pi_N( \beta, \alpha) \theta( \beta ) q_{\textrm{val}}( \beta, t, \mu^2 ) \delta( x - \beta - \alpha \xi ).
\end{equation}
The $t$-dependent PDF $q_{\textrm{val}}( \beta, t, \mu^2 )$ is parameterized as:
\begin{equation}
\label{eq-GK-tdependent-PDF}
q_{\textrm{val}}( \beta, t, \mu^2 ) = \beta^{-\alpha' t} \beta^{- \delta}( 1 - \beta )^{2 n + 1}\sum_{j = 0}^2 c_{j} \beta^{\frac{j}{2}}.
\end{equation}
The coefficients $\delta$ and $c_j$ have been determined in a fit to the CTEQ6m PDFs \cite{Pumplin:2002vw} with $n = 1$. This specific choice of the PDF parameterization allows the analytic computation of the GPD $H^q_{\textrm{val}}$. In practice, we computed numerically  the GPD $H^q_{\text{val}}$ and checked that the analytic calculation gave the same result. The coefficient $\alpha'$ has been chosen in order to approximate the small $t$-dependence of the quark contribution $F^q_1$ to the proton form factor $F_1$. The values of these coefficients for $u$ and $d$ quarks are recalled in \reftab{tab-valence-PDF-coefficients}. 

In \refcite{Goloskokov:2005sd, Goloskokov:2007nt, Goloskokov:2009ia, Kroll:2012sm} the profile function exponent $N$ is fixed to 1, but in the following we will allow it to take real values between 1 and $+\infty$. Such a change of $N$ does not modify the description of the form factor $F_1$:
\begin{eqnarray}
F^q_1( t ) & = & \int_{-1}^{+1} \mathrm{d}x \, \int_\Omega \mathrm{d}\beta\mathrm{d}\alpha \, \pi_N( \beta, \alpha) \theta( \beta ) q_{\textrm{val}}( \beta, t, \mu^2 ) \delta( x - \beta - \alpha \xi ) \nonumber \\
& = & \int_0^{+1} \mathrm{d}\beta q_{\textrm{val}}( \beta, t, \mu^2 ) \int_{-1 + \beta}^{+1 - \beta} \mathrm{d}\alpha \, \pi_N( \beta, \alpha ) \label{eq-formfactor-1CDD-F1-unintegrated} \\
& = & \int_0^{+1} \mathrm{d}\beta q_{\textrm{val}}( \beta, t, \mu^2 ). \label{eq-formfactor-1CDD-F1-integrated} 
\end{eqnarray}
In this derivation we used the normalization (\ref{eq-profile-function-normalization}) between \refeq{eq-formfactor-1CDD-F1-unintegrated} and \refeq{eq-formfactor-1CDD-F1-integrated}.

\begin{table}[h]
\begin{center}
  \begin{tabular}{|c|c|c|}
    \hline
    \hline
    & $u_{\textrm{val}}$ & $d_{\textrm{val}}$ \\
    \hline
    $\delta$                                    & 0.48                       	& 0.48                     \\
    $c_0$                                        & 1.52 + 0.248 L       	& 0.76 + 0.248 L      \\
    $c_1$                                        & 2.88 - 0.940 L     		& 3.11 - 1.36 L       \\
    $c_2$                                        & -0.095 L                	& -3.99 + 1.15 L     \\
    \hline
    $\alpha' (\GeV^{-2})$                & 0.9                         	& 0.9 \\
    $n$					      & 1.				 	& 1. \\
    \hline
    \hline
  \end{tabular}
  \caption{parameterization of the valence quarks PDFs in \refeq{eq-GK-tdependent-PDF} with $L = \log (Q^2 / Q_0^2)$ and $Q_0^2 = 4~\GeV^2$.}
  \label{tab-valence-PDF-coefficients}
\end{center}
\end{table}

Using the following notation:
\begin{equation}
\beta_1 = \frac{x - \xi}{1 - \xi} \quad \textrm{ and } \quad 
\beta_2 = \frac{x + \xi}{1 + \xi}, \label{eq-def-beta1-beta2}
\end{equation}
\refeq{eq-GK-Hval-parameterization} can be further simplified:
\begin{eqnarray}
H_{\textrm{val}}^q( x, \xi, t, \mu^2 ) & = &\frac{1}{\xi} \theta( x > \xi ) \int_{\beta_1}^{\beta_2} \mathrm{d}\beta \, \pi_N\left( \beta, \frac{x - \beta}{\xi} \right) q_{\textrm{val}}( \beta, t, \mu^2 ) \nonumber \\
& & \quad + \frac{1}{\xi} \theta( x < \xi ) \int_{0}^{\beta_2} \mathrm{d}\beta \, \pi_N\left( \beta, \frac{x - \beta}{\xi} \right) q_{\textrm{val}}( \beta, t, \mu^2 ). 
\label{eq-GK-val-nodiracdelta}
\end{eqnarray}
The dependence of the PDF on the factorization scale $\mu^2$ is approximated through the $L$-dependence of the PDF coefficients exhibited in \reftab{tab-valence-PDF-coefficients}. In the GK model, this is the only dependence of the GPDs on the factorization scale. The factorization scale is chosen to be equal to the photon virtuality : $\mu^2 = Q^2$.

\subsection{Modification of the valence sector}
\label{sec:valence-sector-modification}

In \refcite{Radyushkin:2011dh} Radyushkin used a nucleon-inspired PDF toy model $q( \beta ) = ( 1 - \beta )^3 / \sqrt{\beta}$ in the 1CDD framework. The resulting GPD is peaked at $x \simeq \xi$, the height of this peak being a decreasing function of $N$. This means that the imaginary part of the Compton Form Factor $\mathcal{H}^q$ computed at Leading Order (LO) in QCD perturbation theory (\ie $\ImH^{\textrm{LO}}( \xi ) = \pi H^{q+}( \xi, \xi )$) may depend markedly on the profile function exponent $N$. This is of direct phenomenological relevance since measurements in the valence region with a polarized beam directly depends of this quantity \cite{Belitsky:2001ns, Belitsky:2008bz, Belitsky:2010jw}.

Following \refcite{Radyushkin:2011dh}, the analog in the 1CDD framework of the 2CDD expression \refeq{eq-GK-val-nodiracdelta} is:
\begin{eqnarray}
 H_{\textrm{val}}^q( |x| < \xi, \xi, t, \mu^2  ) 
& = & \frac{x}{\xi}\int_0^{\beta_2} \mathrm{d}\beta \, \frac{q_{\textrm{val}}( \beta, t, \mu^2 )}{\beta} \left[ \pi_N\left( \beta, \frac{x - \beta}{\xi} \right) - \pi_N\left( \beta, \frac{x}{\xi}(1-\beta) \right) \right] \nonumber \\
&  & + \frac{x}{\xi}\int_0^{\beta_2} \mathrm{d}\beta \, \frac{q_{\textrm{val}}( \beta, t, \mu^2 )}{\beta} \pi_N\left( \beta, \frac{x}{\xi} ( 1 - \beta ) \right) \left[ 1 - \frac{1}{1-\beta} \right] \nonumber \\
&  & - \frac{x}{\xi}\int_{\beta_2}^{1} \mathrm{d}\beta \, \frac{q_{\textrm{val}}( \beta, t, \mu^2 )}{\beta (1-\beta)} \pi_N\left( \beta, \frac{x}{\xi} ( 1 - \beta )  \right) \label{eq-H-1CDD-corrected-mistake-x-small} \\
H_{\textrm{val}}^q( x > \xi, \xi, t, \mu^2 ) & = & \frac{x}{\xi}\int_{\beta_1}^{\beta_2} \mathrm{d}\beta \, \frac{q_{\textrm{val}}( \beta, t, \mu^2 )}{\beta} \pi_N\left( \beta, \frac{x - \beta}{\xi} \right). \label{eq-H-1CDD-corrected-mistake-x-large} 
\end{eqnarray}
Starting from \refeq{eq-analogous-eq-48-radyushkin-paper} the $F_1$ form factor sum rule writes:
\begin{eqnarray}
F^q_1( t ) & = & \int_{-1}^{+1} \mathrm{d}x \, x \int_\Omega \mathrm{d}\beta\mathrm{d}\alpha \, \pi_N( \beta, \alpha) \theta( \beta ) \frac{q_{\textrm{val}}( \beta, t, \mu^2 )}{\beta} \nonumber \\
&  & \ \times \left[ \delta( x - \beta - \alpha \xi ) - \frac{1}{( 1 - \beta )^2} \delta\left( x - \frac{\alpha \xi}{1 - \beta} \right) \right] \nonumber \\
& = & \int_0^{+1} \mathrm{d}\beta q_{\textrm{val}}( \beta, t, \mu^2 ) \int_{-1 + \beta}^{+1 - \beta} \mathrm{d}\alpha \, \pi_N( \beta, \alpha ) \nonumber \\
& & \quad + \xi \int_0^{+1} \mathrm{d}\beta \, \frac{q_{\textrm{val}}( \beta, t, \mu^2 )}{\beta} \left[1 - \frac{1}{( 1 - \beta )^3} \right] \int_{-1 + \beta}^{+1 - \beta} \mathrm{d}\alpha \, \alpha \pi_N( \beta, \alpha ) \label{eq-formfactor-2CDD-F1-unintegrated} \\
& = & \int_0^{+1} \mathrm{d}\beta \, q_{\textrm{val}}( \beta, t, \mu^2 ). \label{eq-formfactor-2CDD-F1-integrated} 
\end{eqnarray}
In this derivation, we used the normalization (\ref{eq-profile-function-normalization}) and the fact that $\pi_N$ is an even function of $\alpha$ between \refeq{eq-formfactor-2CDD-F1-unintegrated} and \refeq{eq-formfactor-2CDD-F1-integrated}. The lessons of this computation are twofold: firstly, the result \refeq{eq-formfactor-2CDD-F1-integrated} is the same as \refeq{eq-formfactor-1CDD-F1-integrated}. This means that we can change the DD formalism from 1CDD to 2CDD without spoiling the $t$-dependence tuned to fulfill the form factor sum rule. Secondly, the final expression of the sum rule in \refeq{eq-formfactor-2CDD-F1-integrated} does not depend on the exponent $N$ of the profile function. Thus we can also change $N$ without altering this sum rule. Then to observe the phenomenological consequences of the choice of the DD formalism, we just have to change \refeq{eq-GK-val-nodiracdelta} to \refeq{eq-H-1CDD-corrected-mistake-x-small} or \refeq{eq-H-1CDD-corrected-mistake-x-large}.

When $N \rightarrow +\infty$ the profile function $\pi_N( \beta, \alpha ) \rightarrow \delta( \alpha )$. Taking this limit in the classical 2CDD formalism, \refeq{eq-GK-Hval-parameterization} yields:
\begin{equation}
H_{\textrm{val}}^q( x, \xi, t, \mu^2 ) \underset{N \infty}{\longrightarrow} \theta( x ) q_{\textrm{val}}( x, t, \mu^2 ).
\end{equation}
Following the terminology of \refcite{Belitsky:1999sg}, we will quote the resulting asymptotic GPD model \emph{Forward Parton Density} (FPD). The same exercise applied to the 1CDD model \refeq{eq-analogous-eq-48-radyushkin-paper} also gives: 
\begin{equation}
H_{\textrm{val}}^q( x, \xi, t, \mu^2 ) \underset{N \infty}{\longrightarrow} \theta( x ) \frac{q_{\textrm{val}}( x, t, \mu^2 )}{x},
\end{equation} 
where we used $x \delta( x ) = 0$. When $x \neq 0$ both 1CDD and 2CDD models have the \emph{same} limit at large $N$. The resulting GPDs do not depend on the skewness $\xi$.

\begin{figure}[h]
  \centering
  \includegraphics[width = 0.65\textwidth]{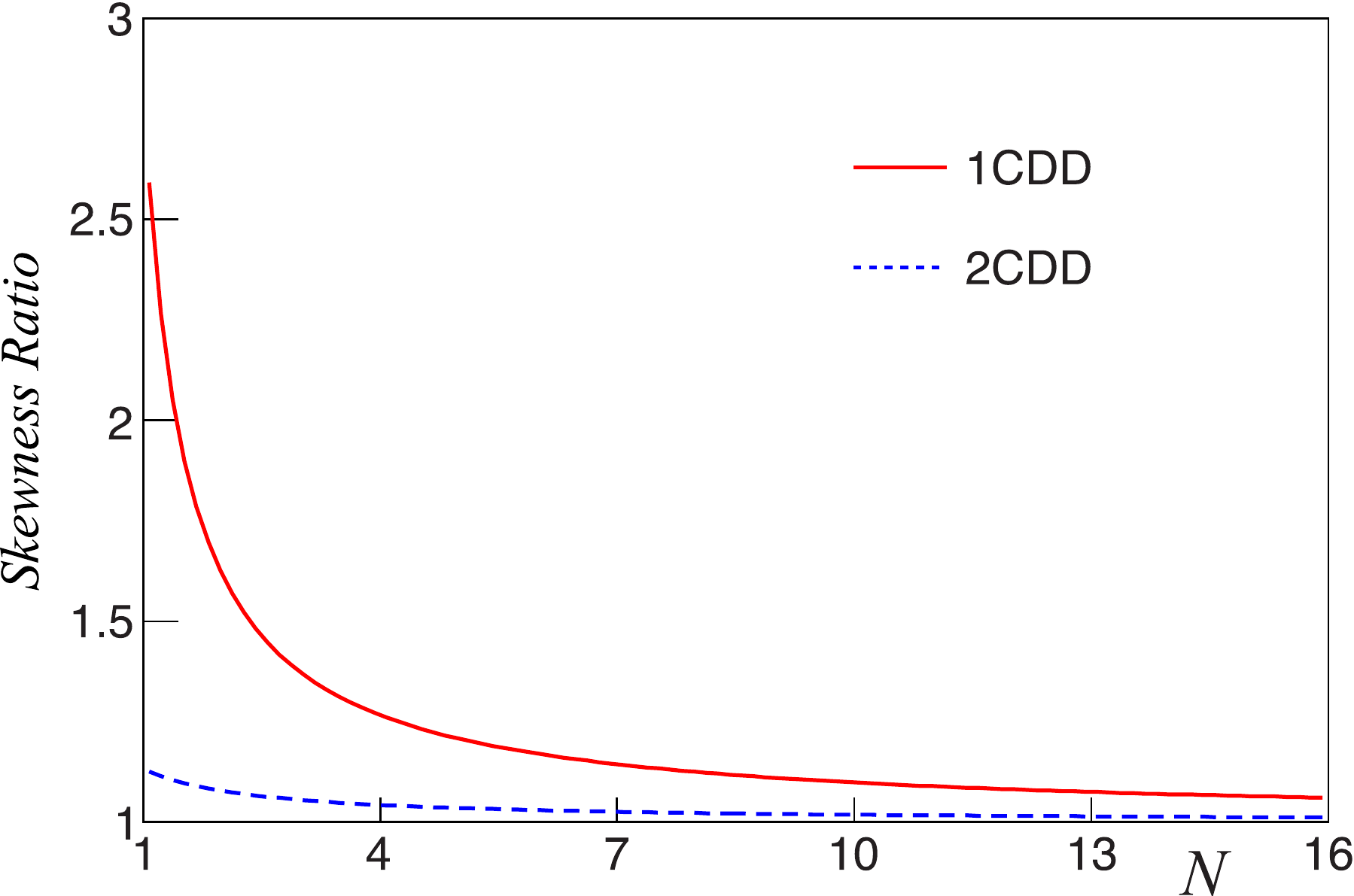}
  \caption{Skewness ratio at the arbitrary value $\alpha=0.5$ as a function of the profile function parameter $N$ for the 1CDD (solid red line) and 2CDD (dashed blue line) parameterizations.}
  \label{fig:SkewnessRatio}
\end{figure}

Consequently, one may ask how the skewness ratio \ie $H(x,x)/H(x,0)$ at small $x$ is modified. Using a Regge-behaved PDF toy model $q(x) \propto x^{-\alpha}$, one can compute this ratio analytically in the small $x$ limit. It has been done for example in \refcite{Kumericki:2009uq} for the 2CDD formalism. We computed this ratio in the case of 1CDD formalism for  $0 < \alpha < 1$ because we used the once-subtracted dispersion relation \refeq{eq-once-subtracted-dispersion-relation}. Both results are different:
\begin{eqnarray}
  \label{eq:SkewnessRatio}
  \frac{H^q( x, x )}{H^q( x, 0 )}\bigg|_{2\text{CDD}} & = & 2^{-\alpha}\ \frac{\Gamma(2N+2)~}{\Gamma(N+1)~}\ \frac{\Gamma(N+1-\alpha)}{\Gamma(2N+2-\alpha)}, \\
  \frac{H^q( x, x )}{H^q( x, 0 )}\bigg|_{1\text{CDD}} & = & 2^{-\alpha}\ \frac{\Gamma(2N+2)~}{\Gamma(N+1)~}\ \frac{\Gamma(N+1-\alpha)}{\Gamma(2N+2-\alpha)}\ \frac{2N+1-\alpha}{2(N-\alpha)} \\
  & = & \frac{H^q( x, x )}{H^q( x, 0 )}\bigg|_{2\text{CDD}}~ \frac{2N+1-\alpha}{2(N-\alpha)}
\end{eqnarray}
The evaluations of the skewness ratio go to 1 as $N$ grows to infinity. On \reffig{fig:SkewnessRatio} we compare the dependence on $N$ of those two ratios at an arbitrary chosen value of $\alpha \in ] 0, 1 [$. When both $\alpha$ and $N$ go to $1$, the skewness ratio is divergent in the 1CDD case.

\section{Comparison to experimental data}

As stated before we will apply this 1CDD GPD model to DVCS measurements in the valence region. We work with a Leading-Order (LO) definition of the CFF $\mathcal{H}^q$:
\begin{eqnarray}
\mathcal{H}^q( \xi, t )
& = & \int_{-1}^{+1} \mathrm{d}x \, H^q( x, \xi, t ) \left( \frac{1}{\xi - x -i \epsilon} - \frac{1}{\xi + x - i \epsilon} \right), \label{eq-def-CFF-H} \\
\ReH^q( \xi, t )
& = & \mathcal{P} \int_{-1}^{+1} \mathrm{d}x \, H^q( x, \xi, t ) \left( \frac{1}{\xi - x} - \frac{1}{\xi + x} \right), \label{eq-def-ReH} \\
\ImH^q( \xi, t )
& = & \pi \big( H^q( x, \xi, t ) - H^q( -x, \xi, t ) \big). \label{eq-def-ImH}
\end{eqnarray}
where $\mathcal{P}$ denotes Cauchy's principal value prescription. We denote $\mathcal{H}$ the average of $\mathcal{H}^q$s weighted by the square of quark electric charges.

It is commonly believed that Next-to-Leading Order (NLO) corrections have a small impact in the valence region, hence justifying the LO approximation. However complete next-to-leading order expressions of CFFs are available \cite{Ji:1998xh, Belitsky:1997rh,  Ji:1997nk, Mankiewicz:1997bk, Belitsky:1999sg, Freund:2001rk, Freund:2001hm, Freund:2001hd, Pire:2011st} and recent estimates \cite{Moutarde:2013qs} challenge the aforementioned common view: quark and gluon NLO contributions may not be negligible even in the valence region at moderate energy. These results triggered an ongoing theoretical effort on the soft-collinear resummation in DVCS \cite{Altinoluk:2012nt, Altinoluk:2012fb}. New developments in this direction are expected in the near future. However our concern here is not a detailed phenomenological study of DVCS in the valence region, but rather the test of a new parameterization of the GPD $H$. Therefore we will stay close to the prescriptions used in recent evaluations of DVCS observables \cite{Kroll:2012sm} and use the LO expressions \refeq{eq-def-ReH} and \refeq{eq-def-ImH} of the CFF $H$.

As stated in the introduction, we will apply the 1CDD formalism to the computation of JLab Hall~A and CLAS polarized beam observables, namely beam helicity dependent and independent cross sections \cite{Munoz Camacho:2006hx} and beam-spin asymmetries \cite{Girod:2007aa}. All details about the evaluation of observables related to the $e p \rightarrow e p \gamma$ channel are given in \refcite{Kroll:2012sm}. In particular we use the Trento convention \cite{Bacchetta:2004jz} for the definition of the angle between the hadronic and leptonic planes. The variable $\xb$ is classically defined as:
\begin{equation}
\xb = - \frac{q^2}{2 p \cdot q},
\label{eq:def-bjorken-x}
\end{equation}
where $p$ and $q$ are the 4-momenta of the target nucleon and the exchanged virtual photon in the Born approximation. We also take the following asymptotic formula for $\xi$:
\begin{equation}
  \xi = \frac{\xb}{2-\xb}.
\end{equation}

\begin{figure}[h]
  \centering
  \includegraphics[width = 0.5\textwidth]{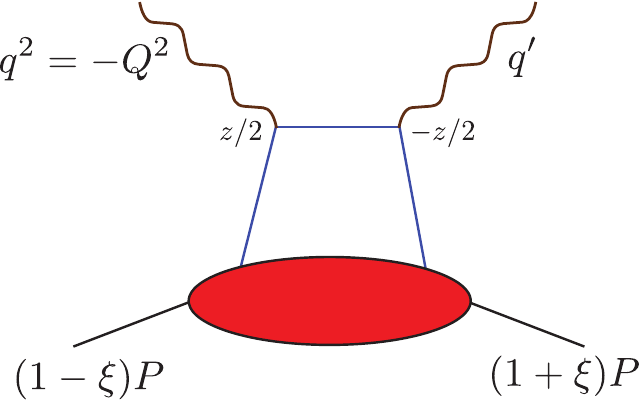}
  \caption{Handbag diagrams for the DVCS process.}
  \label{fig:DVCS}
\end{figure}

When comparing our model with the data, one should keep in mind  every limitation it contains, including the leading order in $\alpha_s$ approximation, the modification of the valence sector alone, the neglect of higher twist corrections or the treatment of QCD evolution equations.

\subsection{Tuning of the profile function}

In the 2CDD formalism it is common to set the exponent $N=1$ in the profile function $\pi_N$ because it corresponds to the asymptotic shape of a non-singlet quark distribution amplitude; this is the choice made in the GK and VGG models. However here we study DVCS data with typical $Q^2$ values between 1 and 4~\GeV$^2$, and $x_B$ between 0.1 and 0.5. It is not clear at all that the asymptotic shape is reached. It is thus interesting to compute the CFF $\mathcal{H}$ and DVCS observables for different values of the profile function exponent $N$. 

In a first step, we fit our GPD parameterization \refeq{eq-H-1CDD-corrected-mistake-x-small} and \refeq{eq-H-1CDD-corrected-mistake-x-large} to Hall~A beam helicity-dependent and independent cross section data \cite{Munoz Camacho:2006hx}. Although measured on a restricted kinematic domain, these data are highly accurate, and as emphasized for example in \refcite{Kroll:2012sm}, the helicity-dependent cross sections are mostly sensitive to the singlet GPD $H^+( \xi, \xi, t, Q^2)$. This offers the possibility to set the width of the profile function through the parameter $N$, and independently to quantify the impact of an additional subtraction term such as $D_0$ in \refeq{eq-additional-DTerm-D0}.

\begin{figure}[h]
	\begin{center}
          \includegraphics[width=0.9\textwidth]{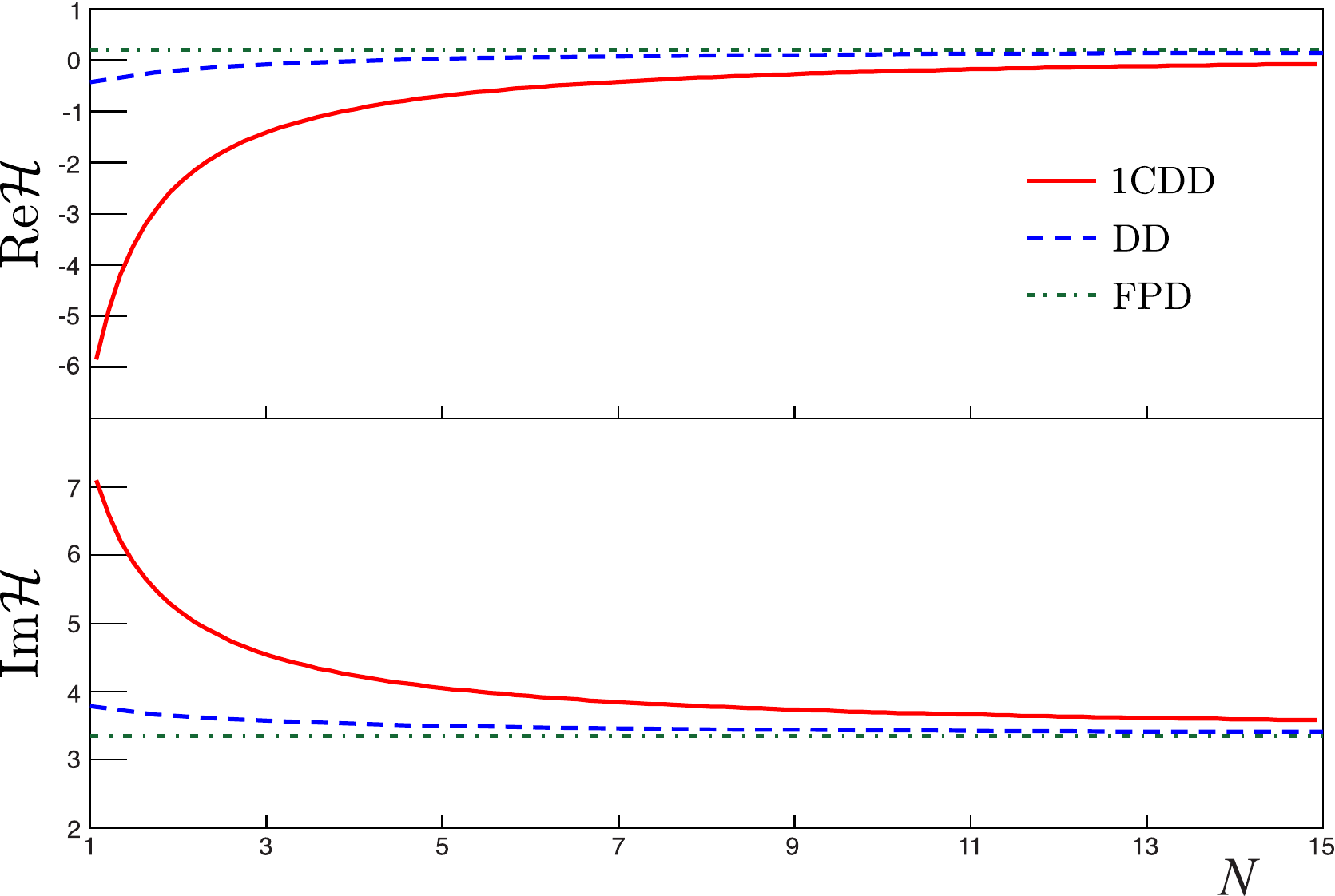}
          \caption{\label{fig-CFF-H-vs-N}CFF $\mathcal{H}$ as a function of the exponent $N$ of the profile function $\pi_N$ at $\xb$ = 0.36, $Q^2$ = 2.3~\GeV$^2$ and $t$ = -0.23~\GeV$^2$ for the 1CDD (solid red line), DD (dashed blue line) and FPD (dash-dotted green line) parameterizations.}
	\end{center}
\end{figure}

First we compute the CFF $\mathcal{H}$ obtained in the 1CDD and 2CDD pictures at the given kinematic configuration $\xb$ = 0.36, $Q^2$ = 2.3~\GeV$^2$ and $t$ = -0.23~\GeV$^2$, and compare it to the CFF obtained in the $N\rightarrow +\infty$ limit (FPD model). We were able to compute CFFs with $N$ ranging from 1 to 50 and managed to get the correct asymptotic behavior, which is a test of our numerical integration routines. On \reffig{fig-CFF-H-vs-N} we observe that the 2CDD model with $N$ = 1 is almost equal to its asymptotic limit. This fact has been known for some time \cite{Goeke:2001tz}. However, we see a marked difference with the 1CDD implementation for small values of $N$. Therefore, we expect a clear difference when computing actual observables. We also expect that for large enough $N$, the 1CDD parameterization will produce similar results to the 2CDD formalism since both have the same asymptotic limit.

\begin{table}[h]
  \centering
  \begin{tabular}{|l|c|}
    \hline
    \hline
    Parameterization 				& $\chi^2 / \text{d.o.f}$ \\
    \hline
    1CCD ($N \simeq 1.86$) 		& 4.0\\
    DD 						& 5.9\\
    FPD 						& 7.2\\
    \hline
    \hline
  \end{tabular}
  \caption{\label{tab:chi-square-HallA}$\chi^2$ per degrees of freedom for the comparison of the different parameterizations with the subset of Hall~A beam helicity-dependent and independent cross sections such that $| t | / Q^2 < 0.1$. No fit was made for the DD or FPD cases, while the parameter $N$ of the profile function was extracted from data in the 1CDD case.}
\end{table}

\begin{figure}[h!]
	\begin{center}
      		\rotatebox{0}{\includegraphics[width=\textwidth]{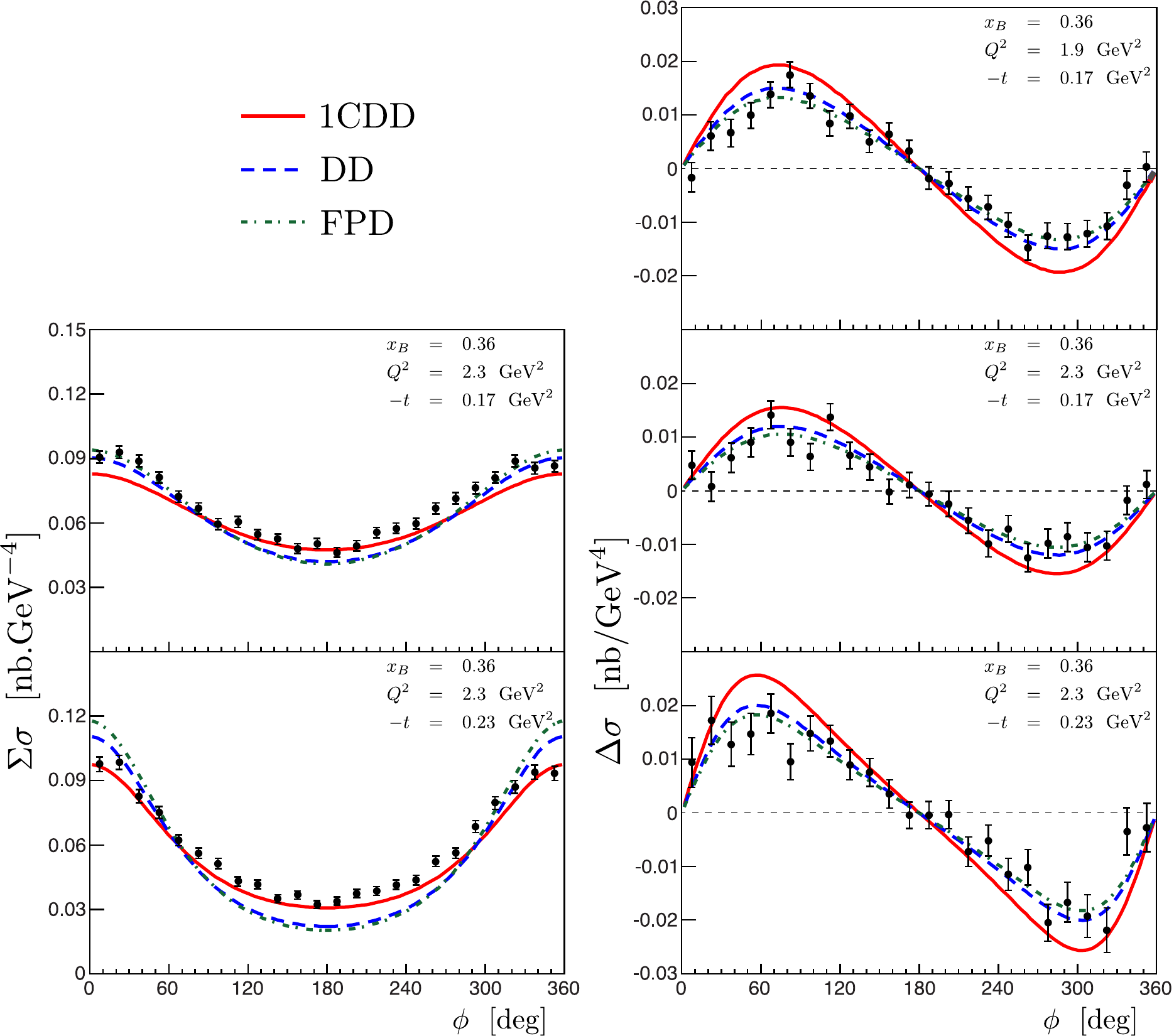}}
  		\caption{\label{fig-HallA}Comparison to Hall~A data : results at $x_B=0.36$, $Q^2=2.3~\GeV^2$ (four lower plots) and  $Q^2=1.9~\GeV^2$ (upper plot), $t = -0.23~\GeV^2$ (two lower plot) and $t = -0.17~\GeV^2$ (three upper plots). The full red line corresponds to the 1CDD model, the dashed blue line to the classical DD Ansatz and the dash-dotted green line to the unskewed FPD limit.}
	\end{center}
\end{figure}

Hall~A helicity-dependent and independent cross sections, restricted on kinematics such that $|t|/Q^2 \le 0.1~$, are best described by 1CDD choosing $N \simeq 1.86$ (see \reftab{tab:chi-square-HallA}). The comparison to part of these data is shown on \reffig{fig-HallA}. In this comparison, all GPDs ($H$, $E$, $\tilde{H}$ and $\tilde{E}$) are taken into account in a leading-twist leading-order evaluation along the lines of \refcite{Kroll:2012sm}. In particular the valence and sea parts of the GPD $H$ are computed, but only the valence part has been changed from the 2CDD to the 1CDD formalism. We observe a discrepancy between the 1CDD prescription and beam helicity-dependent cross sections. This is due to the fact that these data are consistent with the FPD model which is formally accessed in the 1CDD formalism when $N \rightarrow \infty$. From \reffig{fig-CFF-H-vs-N} we see that the 1CDD CFF $H$ at $N \simeq 5$ is already a good approximation of the FPD CFF $H$. Moreover, the smaller $N$ is, the larger $\ImH$ becomes in the 1CDD picture as displayed above in \reffig{fig-CFF-H-vs-N}. On the contrary, this 1-parameter fit leads to a much better agreement with helicity-independent cross sections for $\phi$ close to 0$^\circ$ or 360$^\circ$ and  $\phi$ close to 180$^\circ$ as well.

\subsection{Additional $t$-dependent $D_0$-term}

On the kinematic bin $t=-0.23~\GeV^2$ the improvement brought by the 1CDD picture was significant, whereas it was less satisfactory on the bin $t=-0.17~\GeV^2$. It is still possible to add the subtraction term $D_0$ of \refeq{eq-additional-DTerm-D0}. This term possesses the following distinctive properties:
\begin{description}
\item[(i)\hphantom{ii}] It depends on $t$ through the multiplicative constant $T_0^q$,
\item[(ii)\hphantom{i}] Its contribution $\mathcal{H}_{D_0}^q$ to the CFF $\mathcal{H}^q$ does not depend on $\xi$:
\begin{eqnarray}
Im \mathcal{H}_{D_0}^q( \xi, t ) \label{eq:contribution-S0-to-ImH}
& = & 
0, \label{eq:contribution-S0-to-ImH2} \\
Re \mathcal{H}_{D_0}^q( \xi, t )
& = & \frac{\sqrt{\pi}}{2^{1+2N}} \frac{\Gamma( N )}{N \Gamma\left( \frac{3}{2} + N \right)} T_0^q( t ). \label{eq:contribution-S0-to-ReH}
\end{eqnarray}
\item[(iii)] It contributes only to the highest exponent of the polynomiality relation:
\begin{equation}
\int_{-1}^{+1} \mathrm{d}x \, x^n D_0^q( x, \xi, t ) = T_0^q( t ) \frac{1 + (-1)^{n+1}}{2^{2(1+N)}} \frac{\Gamma\left( 1 + \frac{n}{2} \right) \Gamma( N )}{\Gamma\left( 2 + \frac{n}{2} + N \right)} \xi^{n+1}
\end{equation}
\end{description}
These are the properties of the $D$-term in the classical 2CDD formalism. 

Let us remind that the nucleon $D$-term is not fixed by QCD first principles. It is however customary to define a flavor-singlet $D$-term $D$:
\begin{equation}
D( \alpha, t ) = \sum_{q = u, d, s} D^q( \alpha, t ),
\label{eq:def-flavor-singlet-D-term}
\end{equation}
and to project it onto the basis of Gegenbauer polynomials $C^{3/2}_n$:
\begin{equation}
D( \alpha, t ) = ( 1 - \alpha^2 ) \sum_{n = 0 \atop n \textrm{ odd}}^\infty d_n( t, \mu^2 ) C^{3/2}_n( \alpha ).
\label{eq:def-D-term-projection-Gegenbauer-polynomials}
\end{equation}
The Chiral Quark Soliton Model ($\chi$QSM) gives estimates (see \refcite{Goeke:2001tz} and references therein) of the first three non-vanishing terms of this expansion at a very low scale $\mu_0 \simeq 600~\MeV$ and vanishing momentum transfer:
\begin{eqnarray}
d_1^{u+d}( t = 0~\GeV^2, \mu_0^2 ) & \simeq & - 4.0, \label{eq:ChQSM-d1-low-scale} \\
d_3^{u+d}( t = 0~\GeV^2, \mu_0^2 ) & \simeq & - 1.2, \label{eq:ChQSM-d3-low-scale} \\
d_5^{u+d}( t = 0~\GeV^2, \mu_0^2 ) & \simeq & - 0.4. \label{eq:ChQSM-d5-low-scale} 
\end{eqnarray}
Note that Schweitzer \etal \cite{Schweitzer:2002nm} report a value $d_1^{u+d} \simeq -9.46$ at the low scale $\mu_0$ while Wakamatsu predicts $d_1^{u+d} \simeq - ( 4.9~-~6.2)$ at the same scale for the $\chi$QSM and $d_1^{u+d} \simeq -0.716$ for the MIT Bag model \cite{Wakamatsu:2007uc}. Keeping in mind the overall uncertainty on these parameters, we however evolve the $\chi$QSM coefficients given from \refeq{eq:ChQSM-d1-low-scale} to \refeq{eq:ChQSM-d5-low-scale} perturbatively at LO from the low scale $\mu_0^2$ to the scale $\mu^2 = 2.3~\GeV^2$ of Hall~A measurements. We  follow the treatment of \refcite{Moutarde:2013qs, Berger:2001xd} which assumes a vanishing gluon $D$-term at the low scale $\mu_0^2$ where both quark models are defined:
\begin{eqnarray}
d_1^{u+d}( t = 0~\GeV^2, \mu^2 ) & \simeq & - 3.12, \label{eq:ChQSM-d1-HallA-scale} \\
d_3^{u+d}( t = 0~\GeV^2, \mu^2 ) & \simeq & - 0.71, \label{eq:ChQSM-d3-HallA-scale} \\
d_5^{u+d}( t = 0~\GeV^2, \mu^2 ) & \simeq & - 0.20. \label{eq:ChQSM-d5-HallA-scale} 
\end{eqnarray}

In the specific implementation of the 1CDD formalism we are discussing here, the $D$-term is instead made of two parts. Starting from \refeq{eq-relation-DD-GPD}:
\begin{equation}
D^q( \alpha ) 
= \int_{-1 +| \alpha |}^{+1 - | \alpha |} \mathrm{d}\beta G^q( \beta, \alpha ), \label{eq-def-D-term} 
\end{equation}
we restrict ourselves to the valence contribution to the DD $G$ because the sea contribution to this DD is the original GK model, which is expressed in the DD representation (\ie it is a pure $F$-type DD). Then, from \refeq{eq-G-1CDD} and \refeq{eq:1CDD-structure-szczepaniak-et-al-model} and adding the extra $t$-dependent $D_0$ term (\ref{eq-additional-DTerm-D0-1CDD-picture}):
\begin{eqnarray}
D^q( \alpha )
& = & \alpha \int_{-1 +| \alpha |}^{+1 - | \alpha |} \mathrm{d}\beta \, \big( f^q_{\textrm{val}}( \beta, \alpha ) + f_0^q( \beta, \alpha ) \Big), \nonumber \\
& = & \alpha \int_{-1 +| \alpha |}^{+1 - | \alpha |} \mathrm{d}\beta \, f^q_{\textrm{val}}( \beta, \alpha ) + D_0^q( \alpha ). \label{eq:D-term-with-S0-contribution}
\end{eqnarray}
Using the explicit expression (\ref{eq-analogous-eq-20-radyushkin-proceedings}) of $f^q_{\textrm{val}}( \beta, \alpha )$ we find:
\begin{eqnarray}
D^q( \alpha )
& = & \alpha \int_{-1 +| \alpha |}^{+1 - | \alpha |} \mathrm{d}\beta \, \left[ \frac{q_{\textrm{val}}( \beta )}{\beta} \pi_N( \beta, \alpha ) - \delta( \beta ) \pi_N( 0, \alpha ) \int_0^1 \mathrm{d}\gamma \frac{q_{\textrm{val}}( \gamma )}{\gamma ( 1 - \gamma )^2} \right] + D_0^q( \alpha ),
\end{eqnarray}
and define the 1CDD contribution $D_{\textrm{1CDD}}^q$ to the $D$-term by:
\begin{equation}
D^q( \alpha ) = D_{\textrm{1CDD}}^q( \alpha ) + D^q_0( \alpha ).
\label{eq:def-D-term-1CDD-forward}
\end{equation}
Taking into account the support property (\ref{eq:def-qval-negative-beta}) of $q_{\textrm{val}}$ the previous equation becomes:
\begin{eqnarray}
D^q( \alpha )
& = &  \alpha \int_{0}^{1 - | \alpha |}  \mathrm{d}\beta \, \frac{q_{\textrm{val}}( \beta )}{\beta} \pi_N( \beta, \alpha ) - \alpha \pi_N( 0, \alpha ) \int_0^1 \mathrm{d}\beta \,  \frac{q_{\textrm{val}}( \beta )}{\beta ( 1 - \beta )^2}  + D_0^q( \alpha ) \nonumber \\
& = &  \alpha \int_0^{1 - | \alpha |} \mathrm{d}\beta \, \frac{q_{\textrm{val}}( \beta )}{\beta} \left[ \pi_N( \beta, \alpha ) - \frac{\pi_N( 0, \alpha )}{( 1 - \beta )^2} \right] - \alpha \pi_N( 0, \alpha ) \int_{1 - | \alpha |}^1  \mathrm{d}\beta \, \frac{q_{\textrm{val}}( \beta )}{\beta}\frac{1}{( 1 - \beta )^2}   \nonumber \\
& & \quad + \frac{T_0^q( t )}{2N 2^{2N}} \alpha  ( 1 - \alpha^2 )^N. \label{eq:D-term-1CDD} 
\end{eqnarray}
This $D$-term is indeed made of two very different contributions since only one of them depends on the forward limit $q_{\textrm{val}}$ although both contributions are of course invisible in the forward limit.

\begin{figure}[!h]
	\begin{center}
		\includegraphics[width=0.7\textwidth]{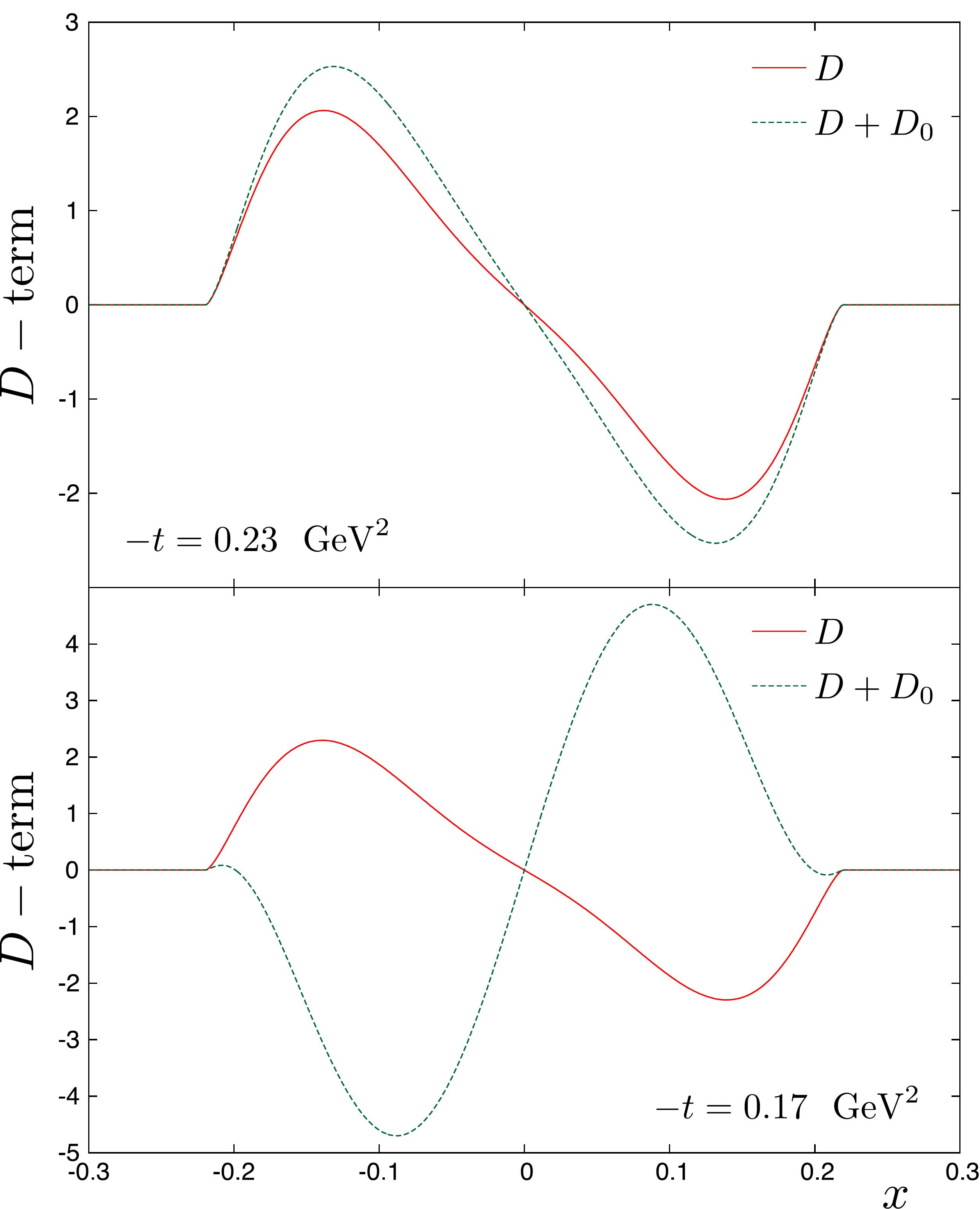}
		\caption{\label{fig:D-term-vs-alpha}$D$-term (\ref{eq:expression-to-compare-to-ChQSM}) vs $x$ with (green dashed curve) and without (full red curve) the $T_0$ contribution for $\xb = 0.36$ ($\xi \simeq 0.22$), $t = -0.17~\GeV^2$ (lower plot) and $t = -0.23~\GeV^2$ (upper plot).}
	\end{center}
\end{figure}

We have seen that adding the $t$-dependent $D_0^q$ term (\ref{eq-additional-DTerm-D0}) only modifies the $D$-term. It will not change the evaluation of helicity-dependent cross sections but may improve the comparison to helicity-independent cross sections. As the $t$-dependence of $T_0^q( t )$ in \refeq{eq:D-term-1CDD} remains unknown, we fit separately its values for the two different Hall~A bins with $| t | / Q^2 < 0.1$. Since we compute the CFFs at LO and with GPDs evaluated at the same scale $\mu^2 = 2.3~\GeV^2$, the fit is only sensitive to the total contribution $D_0( t )$ to the charge and flavor singlet GPD $H^+$:
\begin{equation}
D_0( t ) = 2 \Big( e_u^2 D_0^u( t ) + e_d^2 D_0^q( t ) \Big),
\label{eq:def-charge-flavor-singlet-D-term-contribution}
\end{equation}
where $e_q$ is the quark fractional electric charge (in units of | e |) and the sum is restricted to the lightest flavors because the 1CDD formalism is implemented only in the valence sector. Thus the parameter relevant to the fit is\footnote{We can also obtain this result by  writing the once-subtracted dispersion relation (\ref{eq-once-subtracted-dispersion-relation}) directly with the valence part of $e_u^2 H_u^+ + e_d^2 H_d^+$ and by building the model with the corresponding PDF.}:
\begin{equation}
T_0( t ) = 2 \Big( e_u^2 T_0^u( t ) + e_d^2 T_0^d( t ) \Big).
\label{eq:def-fit-T0}
\end{equation}
At this stage it is reasonable to assume that $T_0^u \simeq T_0^d$ to go further. The flavor-singlet $D$-term we fix from the data is thus:
\begin{eqnarray}
D( \alpha ) & =  & D_{\textrm{1CDD}}^u( \alpha ) + D_{\textrm{1CDD}}^d( \alpha ) + \frac{\Big( T_0^u( t ) + T_0^d( t ) \Big)}{2N 2^{2N}} \alpha  ( 1 - \alpha^2 )^N \nonumber \\
& \simeq & D_{\textrm{1CDD}}^u( \alpha ) + D_{\textrm{1CDD}}^d( \alpha ) + \frac{1}{e_u^2 + e_d^2} \frac{T_0( t )}{2N 2^{2N}} \alpha  ( 1 - \alpha^2 )^N,
\label{eq:expression-to-compare-to-ChQSM}
\end{eqnarray}
and is represented in \reffig{fig:D-term-vs-alpha}. We compare this quantity to the $\chi$QSM estimates in \reftab{tab:comparison-fit-D-term-ChQSM} by projecting \refeq{eq:expression-to-compare-to-ChQSM} onto the basis of orthogonal polynomials $C^{3/2}_n$. Apart from the coefficient $d_3^{u+d}( t = -0.23~\GeV^2 )$ both $\chi$QSM and 1CDD model estimates have the same order of magnitude, the latter being generically by a factor 2 smaller than the former (in absolute value). Remembering the small value of $d_1^{u+d}$ in the MIT Bag model reported by Wakamatsu \cite{Wakamatsu:2007uc}, we conclude that the coefficients $| d_n^{u+d} |$ have the right order of magnitude.

\begin{table}[h]
	\begin{center}
		\begin{tabular}{|c|c|c|c|}
			\hline
			\hline
			~ 	& $\chi$QSM	& \multicolumn{2}{|c|}{Fit} \\
			\cline{2-4} \raisebox{1.5ex}[0pt]{Coefficients}    	& \rule[-2mm]{0mm}{7mm}$t = 0~\GeV^2$	&	$t = -0.17~\GeV^2$	&	 $t = -0.23~\GeV^2$	  \\ \hline
			$d_1^{u+d}$		& - 3.12		& 0.39		& - 1.83 \\
			$d_3^{u+d}$		& - 0.71		& - 0.65		&  0.018 \\
			$d_5^{u+d}$		& - 0.20		& 0.12		&  0.14\\
			\hline
			\hline
		\end{tabular}
		\caption{\label{tab:comparison-fit-D-term-ChQSM}. Comparison of the coefficients of the $D$-term expansion \refeq{eq:def-D-term-projection-Gegenbauer-polynomials} evaluated from the Chiral Quark Soliton Model and extracted from Hall~A data with the Ansatz \refeq{eq:D-term-1CDD}.}
	\end{center}
\end{table}

This coefficient $d_1$ is especially interesting because it can be related to the quark part of the static (\ie defined in the Breit frame) symmetric energy momentum tensor $T_{\mu\nu}^q$ \cite{Polyakov:2002wz}:
\begin{equation}
d_1^q( t = 0~\GeV^2 ) = - \frac{M}{2} \int d^3\vec{r} \, T_{ij}^q( \vec{r} ) \left( r^i r^j - \frac{1}{3} \delta^{ij} r^2 \right),
\label{eq:relation-d1-energy-momentum-tensor}
\end{equation}
where $M$ denotes the nucleon mass and $i$, $j$ are spatial indices. Decomposing $T_{ij}^q$ in a way that makes manifest the distribution of pressure $p( r )$ and shear forces $s( r )$ of the nucleon envisioned as a continuous medium:
\begin{equation}
T_{ij}^q( \vec{r} ) = s^q( r ) \left( r^i r^j - \frac{1}{3} \delta^{ij} r^2 \right) + p^q( r ) \delta_{ij},
\label{eq:def-pressure-shear-distributions}
\end{equation}
one can express $d_1^q$ in terms of $s^q( r )$ or $d^q( r )$ \cite{Goeke:2007fp}:
\begin{equation}
\label{eq:relation-d1-r-p}
d_1^q( t = 0~\GeV^2 ) = - \frac{M}{3} \int d^3\vec{r} \, r^2 s^q(r )  = \frac{5 M}{4} \int d^3\vec{r} \, r^2 p^q(r ).
\end{equation}
In the $\chi$QSM $d_1^{u+d}( t = 0~\GeV^2 )$ has a negative value, and it was conjectured from the above mechanical considerations that it should be the case in general. However Wakamatsu estimated in \refcite{Wakamatsu:2007uc} the valence and sea contributions to the coefficient $d_1^{u+d}$ at a low scale: $d_{1, \textrm{val}}^{u+d} \simeq 0.66$ and $d_{1, \textrm{sea}}^{u+d} \simeq -5.51$. Both sea and valence parts may have opposite signs; this is in agreement with our result which applies only to the valence sector.

\begin{figure}[h!]
	\begin{center}
		\includegraphics[width = 0.7\textwidth]{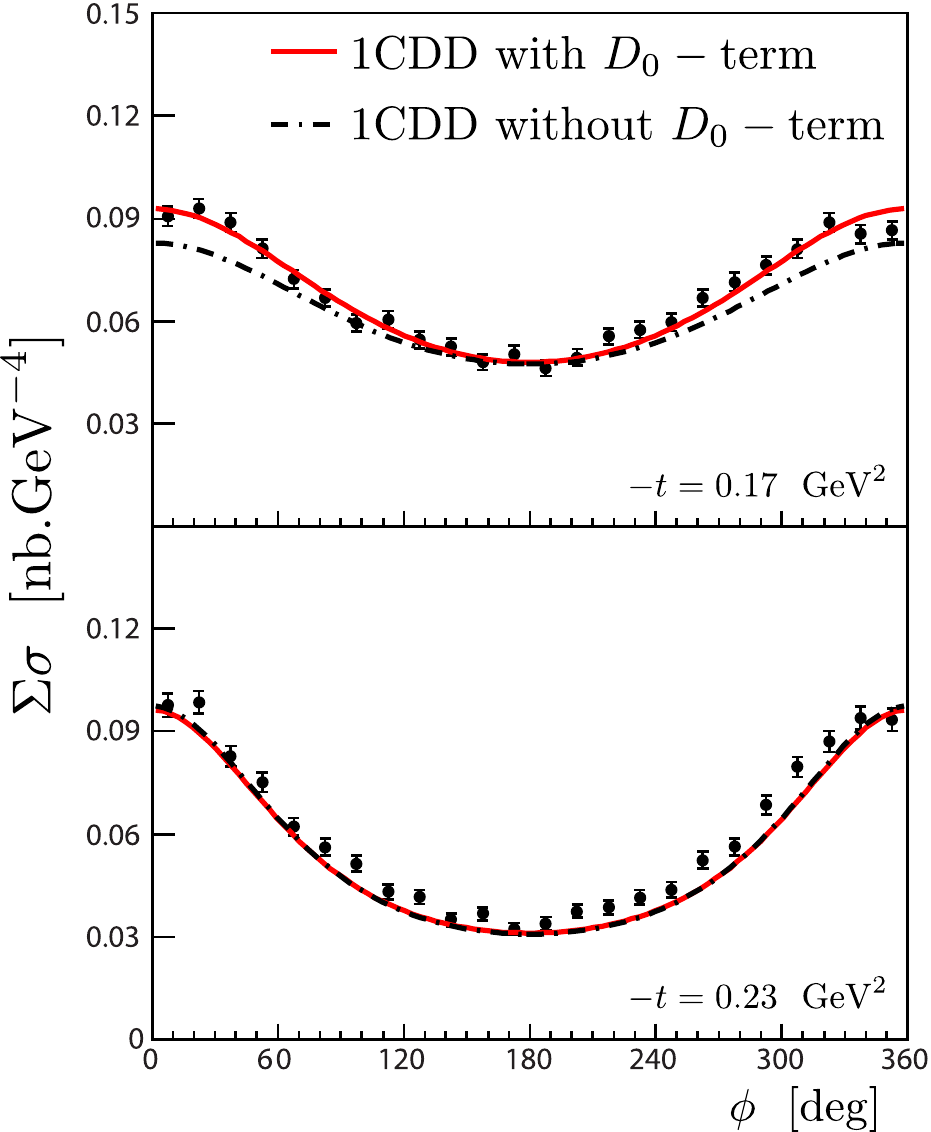}
		\caption{\label{fig-additional-D-term-HallA-cross-sections}Impact of the additional $t$-dependent subtraction term and helicity-independent Hall~A cross sections at $\xb = 0.36$, $Q^2 = 2.3~\GeV^2$, $t = -0.17~ \GeV^2$ (above) and $t = -0.23~ \GeV^2$ (below) for the 1CDD without $D_0$-term (full red line) and with $D_0$-term (dash-dotted black line).}
	\end{center}
\end{figure}

In view of all the approximations involved in our comparison to the considered Hall~A measurements (restriction to leading twist and leading order in perturbation theory, evolution of PDF only, etc.), we will not push further this analysis and now discuss the change in the fit induced by the $t$-dependent $D_0$ term (\ref{eq-additional-DTerm-D0}). It allows to obtain a good agreement to Hall~A helicity-independent cross-sections (see \reffig{fig-additional-D-term-HallA-cross-sections}) for the bin $t=-0.17~\GeV^2$. However the change is not dramatic concerning the bin $t=-0.23~\GeV^2$ and the agreement is still not perfect, even though it is remarkable given the low numbers of fit parameters.

From \refeq{eq:def-D-term-1CDD-forward} we define the contribution $\mathcal{H}^q_D$ to the CFF $\mathcal{H}^q$  associated to the $D$-term:
\begin{equation}
\mathcal{H}^q( \xi, t ) = \mathcal{H}_D^q( \xi, t ) + \mathcal{H}^q_{D_0}( \xi, t ), 
\end{equation}
with:
\begin{equation}
\mathcal{H}_D^q( \xi, t ) = 2 \int_0^{\xi} \mathrm{d}x \, D^q_{\textrm{1CDD}}\left( \frac{x}{\xi} \right) \left( \frac{1}{\xi - x} - \frac{1}{\xi + x} \right),
\label{eq:def-H-D-q}
\end{equation}
and $\mathcal{H}_{D_0}^q( \xi, t )$ is given in \refeq{eq:contribution-S0-to-ImH} and \refeq{eq:contribution-S0-to-ReH}. The contribution $\mathcal{H}_D$ of the $D$-term to the CFF $\mathcal{H}$ is thus:
\begin{equation}
\mathcal{H}_D = e_u ^2 \mathcal{H}_{D}^u( \xi, t ) + e_d ^2 \mathcal{H}_{D}^d( \xi, t ) + \frac{\sqrt{\pi}}{2^{1+2N}} \frac{\Gamma( N )}{N \Gamma\left( \frac{3}{2} + N \right)} \frac{T_0( t )}{2},
\label{eq:contribution-D-term-to-full-CFF}
\end{equation}
since $T_0( t )$ is extracted from a fit of DVCS data via \refeq{eq:def-fit-T0}. The strength of the $D$-term relative to the the real part of the CFF $\mathcal{H}$ is given in \reftab{tab-relative-strength-additional-D-term}. The relative weight of the subtraction term in $\ReH$ suggests that both components are necessary. Moreover a large $D$-term is required, \ie a large $\xi$-independent contribution to $\ReH$. Such a possibility may be tested in the near future thanks to the forthcoming CLAS beam helicity-independent cross sections. Indeed these measurements will offer bins with the same $t$ and $Q^2$ and allow to test the dependence of $\ReH$ on $\xi$.

\begin{table}[h]
	\begin{center}
		\begin{tabular}{|c|c|c|c|c|}
			\hline
			\hline
			$t (\GeV^2)$ 	& $\ReH( T_0 = 0 )$ 	& $\ReH( T_0 \neq 0 )$ 		& $\ReH_D$		& $| \ReH_D / \ReH( T_0 \neq 0 )|$ \\
			\hline
			-0.17 		& -1.59 				& +0.20 					& -0.52				& 2.6  \\
			-0.23 		& -1.65 				& -1.80 					& -2.19				& 1.2 \\
			\hline
			\hline
		\end{tabular}
		\caption{\label{tab-relative-strength-additional-D-term}Evaluation of the real part of the CFF $\mathcal{H}$ with and without the substraction term $D_0$ and impact of the $D$-term (\ref{eq-additional-DTerm-D0}). We remind that $Q^2=2.3~\GeV^2$ and $x_B=0.36$.}
	\end{center}
\end{table}

Even if the improvements induced by a non-vanishing $T_0$ are not sufficient to get a $\chi^2/\textrm{d.o.f.} \approx 1$, our study shows that changing the DD description from 2CDD to 1CDD gives results significantly better than what can be achieved by just adding a $D$-term: here the $\chi^2/\textrm{d.o.f.}$ goes down from $4.0$ to $3.1$ for the set of Hall~A data such that $|t|/Q^2 \le 0.1$. Meanwhile, note that adding a $D$-term such as the one in \refeq{eq-additional-DTerm-D0} to the considered 2CDD parameterization lowers the $\chi^2/\textrm{d.o.f.}$ from $5.9$ to $5.1$ for the same dataset. \reftab{tab:balance-sum-dif-cross-sections} displays the balance between helicity-dependent and helicity-independent cross sections in the fits.

\begin{table}[h]
	\begin{center}
		\begin{tabular}{|c|c|c|c|c|}
			\hline
			\hline
			~ 	& \multicolumn{2}{|c|}{$t = -0.17~\GeV^2$}	& \multicolumn{2}{|c|}{$t = -0.23~\GeV^2$} \\
			\cline{2-5} \raisebox{1.5ex}[0pt]{Ansatz}    	& \rule[-2mm]{0mm}{7mm} $\Sigma\sigma$	& $\Delta\sigma$	& $\Sigma\sigma$	& $\Delta\sigma$ \\ \hline
			1CDD		& 28 / 24		& 158 / 48	& 114 / 24 	& 73 / 24 \\
			DD+D		& 82 / 24		& 50 / 48		&  445 / 24	& 25 / 24 \\
			\hline
			\hline
		\end{tabular}
		\caption{\label{tab:balance-sum-dif-cross-sections}. Statistical weight of helicity-independent ($\Sigma\sigma$) and helicity-dependent ($\Delta\sigma$) cross sections materialized by the $\chi^2$ per data point for both 1CDD and DD+D Ans\"atze. The relative uncertainty on helicity-independent measurements is smaller than the corresponding helicity-dependent datasets, but there are many more data points for helicity-dependent cross sections. We considered Hall~A data such that $| t | / Q^2 < 0.1$.}
	\end{center}
\end{table}

It is presumably possible to improve the agreement with the data. Indeed, the choice of a larger value of $N$ would make the 1CDD model tend to the FPD limit and thus improve the comparison to helicity-dependent cross sections. This would reduce the $\chi^2/\textrm{d.o.f.}$ because there are more data points for helicity-dependent than for helicity-independent cross sections.  Moreover, the loss of agreement on helicity-independent cross sections may be compensated thanks to the additional subtraction term, but it would increase its relative weight in $\ReH$, eventually putting all the GPD physics in the $D$-term. As said before, it will be possible to test this scenario on future datasets. For the time being, we consider that  the other limits of the present phenomenological application presumably precludes from obtaining a fit with a $\chi^2/\textrm{d.o.f.} \simeq 1$. Note also the ability of the 1CDD parameterization to deal with beam helicity-independent cross sections, while it encounters some difficulty with beam helicity-dependent cross sections. Since this 1CDD parameterization was developed for spinless targets, this may suggest that a 1CDD picture adapted to spin-1/2 targets would improve the agreement. This point will be addressed in the last part of this paper.

\subsection{Comparison to CLAS data}

Beam spin asymmetries have been measured in the same kinematic range by the CLAS collaboration \cite{Girod:2007aa}. It is natural to compare the output of the 1CDD implementation adjusted using Hall~A cross sections to CLAS data. Considering those data in the same kinematic range, \ie with $ t \approx -0.17~\GeV^2$ and $|t|/Q^2 \le 0.1$, we compare the evaluations of beam spin asymmetries $\chi^2/\textrm{d.o.f.}$ of the 2CDD and 1CDD models. 

\begin{figure}[h]
  	\begin{center}
		\begin{tabular}{c}
  		\includegraphics[width=\textwidth]{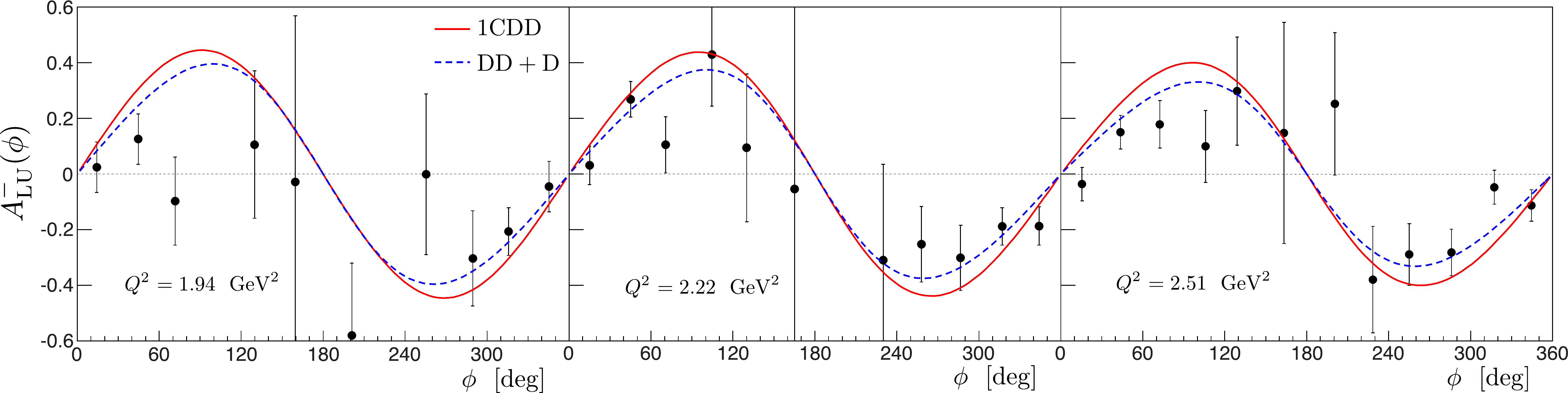} 
		\end{tabular}
  		\caption{Comparison of 1CDD (full red line) and DD+D (dashed blue line) models with CLAS data at $t \simeq -0.17~\GeV^2$ and such that $\frac{|t|}{Q^2} \le 0.1$. From left to right: $\xb = 0.3205$, $t = -0.1705~\GeV^2$ and $Q^2 = 1.9424~\GeV^2$; $\xb = 0.3215$, $t = -0.1719~\GeV^2$ and $Q^2 = 2.217~\GeV^2$; $\xb = 0.3215$, $t = -0.1743~\GeV^2$ and $Q^2 = 2.5078~\GeV^2$.}
  		\label{fig:Comparison_Hall_B}
	\end{center}
\end{figure}

The situation is less favorable for the 1CDD implementation but can easily be explained by the fact that beam spin asymmetries and helicity-dependent cross sections are both mostly sensitive to the imaginary part of the CFF $\mathcal{H}$. In the 1CDD implementation $\ImH$ is larger than in the FPD model, and the FPD model is in good agreement with these data. The clear advantage of the 1CDD parameterization appears in the description of helicity-independent cross sections, in particular near $\phi$ = 180$^\circ$ where helicity-dependent cross sections vanish. All implementations (1CDD, 2CDD and FPD) give similar estimates of helicity-dependent cross sections near $\phi$ = 60 to 90$^\circ$, which is where beam spin asymmetries are maximum. The improvement brought by the 1CDD implementation is then hidden because there are many more data sensitive to $\ImH$ than to $\ReH$. However the flexibility of the 1CDD representation still exists, and we have checked that choosing $N \simeq 5$ produces estimates of Hall~A cross sections similar to those from the 2CDD or FPD frameworks. So even if the data do not show a clear advantage of the 1CDD picture, we do not loose anything by using this representation since it can be tuned to yield a similar quality of agreement with data.


\section{Extension to the 1CDD representation of nucleon GPDs}

\subsection{From the spinless to the spin-1/2 case}

At the time of writing this paper, Radyushkin published an extension of his earlier work on spinless target \cite{Radyushkin:2011dh, Radyushkin:2012gba} to spin-1/2 targets \cite{Radyushkin:2013hca}. The key observation is the following: it is natural to model $H+E$ in the DD representation, and $E$ in the 1CDD formalism. To see why it is so, let us introduce the  twist-2 quark operator:
\begin{equation}
O^q_{\mu\mu_1\ldots\mu_n} = \bar{q}( 0 ) \mathcal{S} \Big( \gamma_\mu i\SymCovDev_{\mu_1} \ldots i\SymCovDev_{\mu_n} \Big) q( 0 ),
\label{eq:def-quark-twist2-operator}
\end{equation}
where $\SymCovDev = ( \RightCovDev - \LeftCovDev ) / 2$ is the covariant derivative and the operator $\mathcal{S}$ projects a tensor onto its completely symmetric and traceless component. The matrix element of this operator between two nucleon states writes:
\begin{eqnarray}
\bra{P+\frac{\Delta}{2}} O^q_{\mu\mu_1\ldots\mu_n} \ket{P-\frac{\Delta}{2}}
& = & \bar{u}\left( P+\frac{\Delta}{2} \right) \left[ \sum_{k=0}^n A_{nk} 
\mathcal{S}\left( \gamma_\mu P_{\mu_1} \ldots P_{\mu_{n-k}}  \frac{-\Delta_{\mu_{n-k+1}}}{2} \ldots \frac{-\Delta_{\mu_n}}{2} \right) \right. \nonumber \\
& & + \sum_{k=0}^n B_{nk} 
\mathcal{S}\left( \frac{i \sigma_{\mu \nu} \Delta^\nu}{2M} P_{\mu_1} \ldots P_{\mu_{n-k}} \frac{-\Delta_{\mu_{n-k+1}}}{2} \ldots \frac{-\Delta_{\mu_n}}{2} \right) \nonumber \\
& &  \left. + \sum_{k=0}^n C_{nk} 
\mathcal{S}\left( \frac{- \Delta_\nu}{2M} P_{\mu_1} \ldots P_{\mu_{n-k}} \frac{-\Delta_{\mu_{n-k+1}}}{2} \ldots \frac{-\Delta_{\mu_n}}{2} \right) \right] u\left( P - \frac{\Delta}{2} \right), \nonumber \\
\label{eq-def-Lorentz-structure-nucleon-target}
\end{eqnarray}
where the coefficients $A_{nk}$, $B_{nk}$ and $C_{nk}$ depend on $t$. We define the DDs $F^q$, $G^q$ and $K^q$ as the generating functions of these coefficients\footnote{We essentially follow the presentation of \refcite{Tiburzi:2004qr} and use the results therein but we correct several typos.}:
\begin{eqnarray}
\frac{n!}{(n-k)! k!} \int_\Omega \mathrm{d}\beta\mathrm{d}\alpha \, \beta^{n-k} \alpha^k F^q( \beta, \alpha ) & = & A_{nk}, \label{eq:def-DD-F-nucleon-target} \\
\frac{n!}{(n-k)! k!} \int_\Omega \mathrm{d}\beta\mathrm{d}\alpha \, \beta^{n-k} \alpha^k K^q( \beta, \alpha ) & = & B_{nk}, \label{eq:def-DD-F-nucleon-target2} \\
\frac{n!}{(n-k)! k!} \int_\Omega \mathrm{d}\beta\mathrm{d}\alpha \, \beta^{n-k} \alpha^k G^q( \beta, \alpha ) & = & C_{nk}. \label{eq:def-DD-F-nucleon-target3} 
\end{eqnarray}
The analogous of \refeq{eq-def-DD-f-spinless-target-1CDD} in the case of a nucleon target is thus:
\begin{eqnarray}
\bra{P+\frac{\Delta}{2}} \bar{q}\left( -\frac{z}{2} \right) \gamma_\mu q \left( \frac{z}{2} \right) \ket{P-\frac{\Delta}{2}}_{z^2=0} 
& = & \bar{u}\left( P+\frac{\Delta}{2} \right) \left[ \gamma_{\mu}\int_{\Omega} \mathrm{d}\beta\mathrm{d}\alpha \, e^{- i \beta (P z) + i \alpha \frac{(\Delta z)}{2}} F^q( \beta, \alpha, t ) \right. \nonumber \\
& & + \frac{i \sigma_{\mu\nu} \Delta^\nu}{2M} \int_{\Omega} \mathrm{d}\beta\mathrm{d}\alpha \, e^{- i \beta (P z) + i \alpha \frac{(\Delta z)}{2}} K^q( \beta, \alpha, t ) \nonumber \\
& & \left. - \frac{\Delta_\mu}{2M} \int_{\Omega} \mathrm{d}\beta\mathrm{d}\alpha \, e^{- i \beta (P z) + i \alpha \frac{(\Delta z)}{2}} G^q( \beta, \alpha, t )\right] u\left( P - \frac{\Delta}{2} \right)  \nonumber \\
& & +\text{ higher twist terms}.
\label{eq-def-DD-F-G--K-nucleon-target}
\end{eqnarray}
The GPDs $H$ and $E$ are defined by:
\begin{eqnarray}
\label{eq-def-GPD-H-E-nucleon-target}
\int \frac{\mathrm{d}z^-}{4\pi} \, e^{i x P^+ z^-} \bra{P+\frac{\Delta}{2}} \bar{q}\left( -\frac{z}{2} \right) \gamma^+ q \left( \frac{z}{2} \right) \ket{P-\frac{\Delta}{2}}_{z^+=0 \atop z_\perp=0} 
& = & \frac{H^q( x, \xi )}{2P^+} \bar{u}\left( P+\frac{\Delta}{2} \right) \gamma^+ u\left( P-\frac{\Delta}{2} \right)  \nonumber \\
& & + E^q( x, \xi ) \bar{u}\left( P+\frac{\Delta}{2} \right) \frac{i \sigma^{+\nu} \Delta_\nu}{2M} u\left( P-\frac{\Delta}{2} \right),  \nonumber \\
& & \label{eq:def-GPD-H-H-nucleon-target}
\end{eqnarray}
which yields the following relations between nucleon DDs and GPDs:
\begin{eqnarray}
H^q( x, \xi ) & = & \int_\Omega \mathrm{d}\beta\mathrm{d}\alpha \, \delta( x - \beta - \alpha \xi ) \Big( F^q( \beta, \alpha ) + \xi G^q( \beta, \alpha ) \Big) \label{eq:relation-GPD-H-DD-F-G}, \\
E^q( x, \xi ) & = & \int_\Omega \mathrm{d}\beta\mathrm{d}\alpha \, \delta( x - \beta - \alpha \xi ) \Big( K^q( \beta, \alpha ) - \xi G^q( \beta, \alpha ) \Big) \label{eq:relation-GPD-E-DD-K-G}.
\end{eqnarray}
It clearly appears that the DD $(F+K)^q$ generates the whole GPD $(H+E)^q$ (\ie no $D$-term is needed in the Polyakov~-~Weiss gauge) and is analogous to an $F$-type DD in the spinless case: this is what we called the DD representation in \refsec{sec:valence-sector-modification}.  Moreover both \refeq{eq:relation-GPD-H-DD-F-G} and \refeq{eq:relation-GPD-E-DD-K-G} are reminiscent of the spinless case \refeq{eq-relation-DD-GPD}. We know from \refsec{sec:1CDD-2CDD-spinless-target} that there exists a gauge in which $H^q$ or $E^q$ can be cast into the 1CDD formalism. Indeed a gauge transformation of nucleon DDs writes\footnote{We refer to \refcite{Tiburzi:2004qr} for a rigorous treatment of the boundary conditions on DDs.}:
\begin{eqnarray}
F^q( \beta, \alpha ) & \rightarrow & F^q( \beta, \alpha ) + \frac{\partial \chi^q}{\partial \alpha}( \beta, \alpha ), \label{eq:gauge-transform-F-nucleon-target} \\ 
K^q( \beta, \alpha ) & \rightarrow & K^q( \beta, \alpha ) - \frac{\partial \chi^q}{\partial \alpha}( \beta, \alpha ), \label{eq:gauge-transform-K-nucleon-target} \\ 
G^q( \beta, \alpha ) & \rightarrow & G^q( \beta, \alpha ) - \frac{\partial \chi^q}{\partial \beta}( \beta, \alpha ), \label{eq:gauge-transform-G-nucleon-target} 
\end{eqnarray}
and $F^q + K^q$ is gauge-invariant. The extension of the 1CDD modeling of the spinless case to the spin-1/2 case proposed by Radyushkin in \refcite{Radyushkin:2013hca} consists in modeling $H+E$ with a single DD and $E$ in the 1CDD framework.

\subsection{The regularizing role of the $D$-term}

Following \refcite{Radyushkin:2011dh, Radyushkin:2013hca}, let us describe the general construction underlying the implementation of the 1CDD representation. In this framework GPD a GPD $F$ generically can be expressed by means of a DD $f$: 
\begin{equation}
\label{eq:generic-1CDD-representation}
F( x, \xi ) = x \int_\Omega \mathrm{d}\beta \mathrm{d}\alpha \,  \delta( x - \beta - \alpha \xi ) f( \beta, \alpha ).
\end{equation}
This DD $f$ generates a $D$-term:
\begin{equation}
D( \alpha ) = \alpha \int_{-1 + | \alpha |}^{+1 - | \alpha |} \mathrm{d}\gamma \, f( \gamma, \alpha ).
\label{eq:def-generated-D-term}
\end{equation}
The complementary part is denoted $[f]_+$:
\begin{equation}
f( \beta, \alpha ) = [f]_+( \beta, \alpha ) + \delta( \beta ) \frac{D( \alpha )}{\alpha}.
\label{eq:generic-decomposition-of-DD}
\end{equation}
The decomposition can be brought at the GPD level:
\begin{eqnarray}
F( x, \xi ) 
& = & [F]_+( x, \xi ) + F_D( x, \xi ), \label{eq:def-plus-part-D-part-GPD} \\
\hphantom{}[F]_+( x, \xi ) 
& = & x \int _\Omega \mathrm{d}\beta \mathrm{d}\alpha \,  \Big( \delta( x - \beta - \alpha \xi ) - \delta( x - \alpha \xi ) \Big) f( \beta, \alpha ), \label{eq:def-plus-part-GPD} \\
F_D( x, \xi ) 
& = & x \int_{-1}^{+1} \mathrm{d}\alpha \, \frac{D( \alpha )}{\alpha} \delta( x - \alpha \xi ). \label{eq:def-D-part-GPD} 
\end{eqnarray}
Let $\phi$ denote a PDF-like function (the forward limit of the GPD $F$) such that $\phi( \beta ) \propto 1/\beta^a$ with $0 < a < 1$. When applying the FDDA to $f( \beta, \alpha )$:
\begin{equation}
f( \beta, \alpha ) = \pi_N( \beta, \alpha ) \frac{\phi( \beta )}{\beta},
\label{eq:generic-FDDA-1CDD}
\end{equation}
the $D$-term contribution to $[F]_+$ in \refeq{eq:def-plus-part-GPD}  guarantees the convergence of the whole integral. Similarly to \refeq{eq-H-1CDD-corrected-mistake-x-small} and \refeq{eq-H-1CDD-corrected-mistake-x-large} we can indeed write for a valence $\phi$ at small $x > 0$:
\begin{eqnarray}
\hphantom{}[F]_+( x, \xi ) 
& = & \frac{x}{\xi} \int_0^{\beta_2} \mathrm{d}\beta \, \left[ \pi_N\left( \beta, \frac{x - \beta}{\xi} \right) - \pi_N\left( \beta, \frac{x}{\xi} \right) \right] \frac{\phi( \beta )}{\beta} \nonumber \\
& & \quad - \int_{\beta_2}^{1 - \frac{x}{\xi}} \mathrm{d}\beta \, \pi_N\left( \beta, \frac{x}{\xi} \right) \frac{\phi( \beta )}{\beta}, \label{eq:D-term-regularization}
\end{eqnarray}
and:
\begin{equation}
\pi_N\left( \beta, \frac{x - \beta}{\xi} \right) - \pi_N\left( \beta, \frac{x}{\xi} \right) = - \frac{N}{4^N x} \left( 1 - \frac{x^2}{\xi^2} \right)^{N-1} \frac{\Gamma( 2 + 2 N )}{\Gamma( 1 + N )^2} \beta + \mathcal{O}( \beta^2 ).
\end{equation} 
This is sufficient to ensure the convergence of the integral in \refeq{eq:def-plus-part-GPD} since $\phi$ is supposed to possess an integrable singularity. The general case can be derived in the same way.

The $D$-term can thus be used to regularize this 1CDD implementation and can be fixed by comparison to experimental or model inspired data. In such a situation this 1CDD implementation of $f$ is defined by the $[f]_+ + D$ prescription which acts as a renormalization prescription.

\subsection{Double Distribution models for $H$ and $E$}

In view of our comparison to DVCS measurements in the valence region, we still modify the valence sector of the GPDs $H^q$ and $E^q$ for $q = u, d$. We still use the GK model as  a basis for our modifications, and describe here the modeling of the GPD $E$ (the GPD $H$ was described in \refsec{sec:H-valence-GK}).

In the original GK model the valence part of the GPD $E$ relies on a forward-like function $e_{\textrm{val}}^q$ and the usual FDDA \ie:
\begin{equation}
\label{eq:GK-Eval-parametrization}
E_{\textrm{val}}^q( x, \xi, t ) = \int_\Omega \mathrm{d}\beta\mathrm{d}\alpha \, \delta( x - \beta - \alpha \xi ) \pi_N( \beta, \alpha) \theta( \beta ) e^q_{\textrm{val}}( \beta, t ),
\end{equation}
with: 
\begin{equation}
\label{eq:GK-tdependent-Forward-Like}
e^q_{\textrm{val}}(\beta, t ) = \beta^{-\alpha' t} \frac{\kappa_q}{B(1 - \mu_{\textrm{val}} , 1 + \nu_{\textrm{val}})} \beta^{-\mu_{\textrm{val}}} (1 - \beta )^{\nu_{\textrm{val}}} ,
\end{equation}
where $B$ is the Euler Beta function. The values of the coefficients of \refeq{eq:GK-tdependent-Forward-Like} are given in \reftab{tab:valence-E-PDF-coefficients}.

\begin{table}[h]
\begin{center}
  \begin{tabular}{|c|c|c|}
    \hline
    \hline
    						& $u$ 		& $d$ \\
    \hline
    $\kappa$     			& 1.67    		&  -2.03   \\
    $\nu_{\textrm{val}}$   		&  4    		&  5.6     \\
    $\mu_{\textrm{val}}$ 	 	&  0.48   		&  0.48   \\
    \hline
    $\alpha' (\GeV^{-2})$   	& 0.9               & 0.9 \\
    \hline
    \hline
  \end{tabular}
  \caption{Parameters of forward-like GPD $E( x, 0, 0 )$ in \refeq{eq:GK-tdependent-Forward-Like}.}
  \label{tab:valence-E-PDF-coefficients}
\end{center}
\end{table}

As announced before, we now describe $H+E$ and $E$ in the DD and 1CDD formalisms supplemented by the FDDA. Let us recall the breaking of the DD gauge invariance after the choice of the gauge where the Ansatz is applied.  Each of the profile functions involved in either $H+E$ or $E$ can come with a different parameter:
\begin{eqnarray}
H_{\textrm{val DD}}^q( x, \xi ) + E_{\textrm{val DD}}^q( x, \xi )
& = & \int_\Omega \mathrm{d}\beta \mathrm{d}\alpha \, \delta( x - \beta - \alpha \xi ) \pi_{N_{H+E}}( \beta, \alpha ) \theta( \beta ) \Big( q_{\textrm{val}}( x ) + e_{\textrm{val}}( x ) \Big), \nonumber \\
& & \label{eq:FDDA-HE-nucleon-target} \\
\hphantom{}[E_{\textrm{val}}^q]_+( x, \xi ) 
& = & x \int_{\Omega} \textrm{d}\beta \textrm{d}\alpha \, \frac{e^q_{\textrm{val}}(\beta)}{\beta} \pi_{N_E} (\beta, \alpha) \Big( \delta( x - \beta - \alpha \xi ) - \delta( x-\alpha \xi ) \Big), \nonumber \\
& & \label{eq:FDDA-E-nucleon-target}
\end{eqnarray}
and the full GPD $E_{\textrm{val}}^q$ is obtained after addition of the $D$-term $D$ introduced to regularize the 1CDD formalism:
\begin{eqnarray}
\label{eq:Nucleon-GPD}
H_{\textrm{val}}^q( x, \xi ) 
& = & H_{\textrm{val DD}}^q( x, \xi ) + E_{\textrm{val DD}}^q( x, \xi ) - [E_{\textrm{val}}^q]_+( x, \xi ) + D\left( \frac{x}{\xi} \right), \label{eq:Hval-nucleon-model} \\
E_{\textrm{val}}^q( x, \xi ) 
& = & [E_{\textrm{val}}^q]_+( x, \xi ) - D\left( \frac{x}{\xi} \right). \label{eq:Eval-nucleon-model}
\end{eqnarray}
We have chosen the following simple functional form for the $D$-term:
\begin{equation}
\label{eq:D-Term-nucleon}
D( \alpha ) = C \alpha ( 1 - \alpha^2 ),
\end{equation}
where $C$ depends on $t$ and is fixed from the data, using this 1CDD implementation as a renormalization prescription.

\begin{figure}[!h]
\begin{center}
\includegraphics[width= 0.7\textwidth]{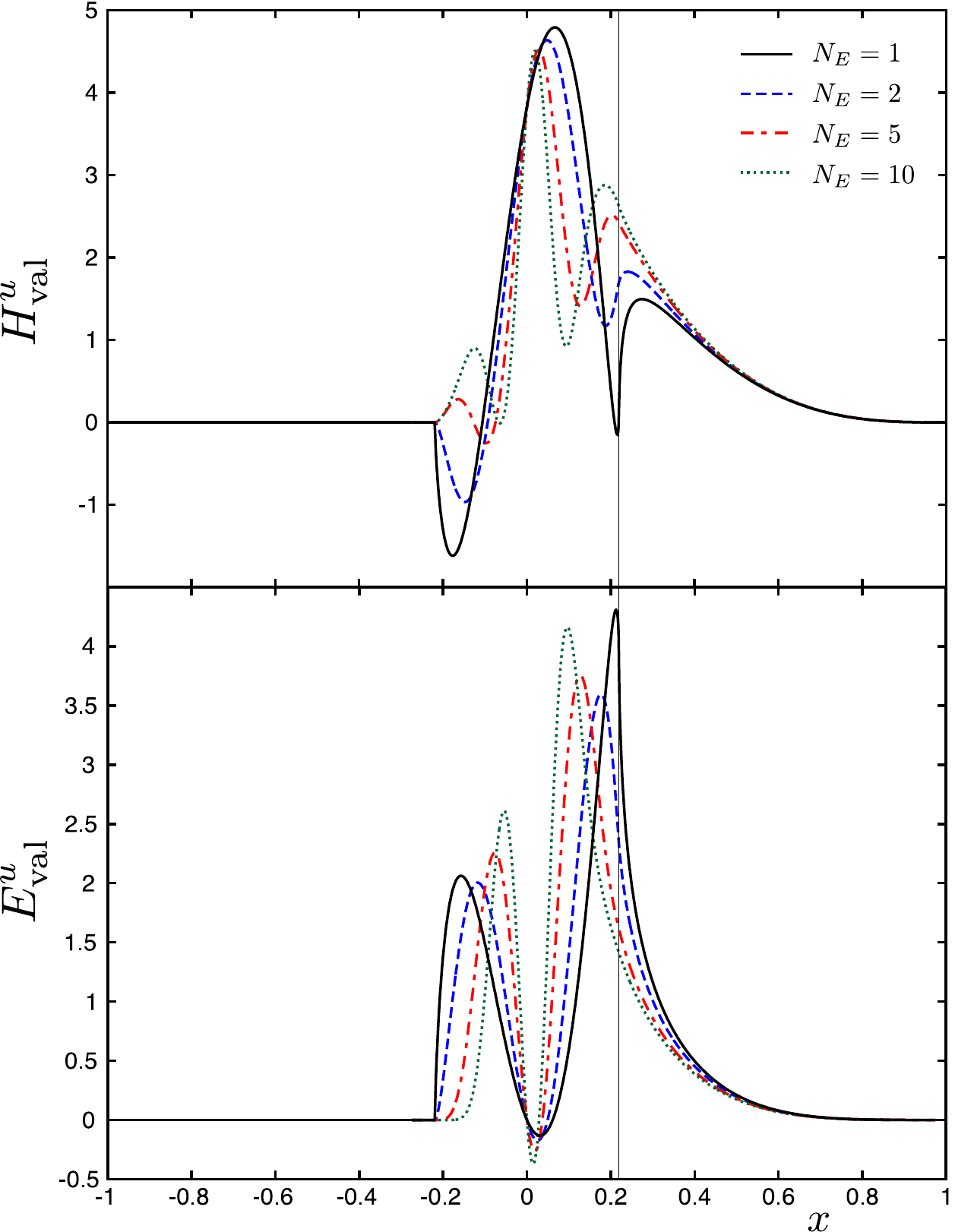}
\caption{\label{fig:GPD-H-E-nucleon-target}Valence contributions to the GPDs $H^u$ in Eq.~(\ref{eq:Hval-nucleon-model}) (upper plot) and $E^u$ in Eq.~(\ref{eq:Eval-nucleon-model}) (lower plot) vs $x$ for different values of $N_E$ in the nucleon model for $\xb = 0.36$, $t = -0.23~\GeV^2$ and $Q^2 = 2.3~\GeV^2$. The vertical black line signal $x = \xi$. In these plots the $D$-term (\ref{eq:D-Term-nucleon}) is arbitrarily set to 0.}
\end{center}
\end{figure}

\reffig{fig:GPD-H-E-nucleon-target} shows the GPDs $H^u_{\textrm{val}}$ and $E^u_{\textrm{val}}$ with $N_{H+E} = 1$ and $N_E$ ranging between 1 and 10.  Since we know that the DD parameterization weakly depends on the profile function parameter, the model is expected to be most sensitive to the exponent of the profile function in the 1CDD sector. The curve corresponding to $N_E = 10$ is close to the FPD limit and gives an indications of the convergence of this family of models when $N_E$ gets large. $H^u_{\textrm{val}}$ has a characteristic doubly-peaked structure which is more complex than its analog in the spinless case. The peaks also get narrower when $N_E$ is increased. The plot of $E^u_{\textrm{val}}$ vs $x$ oscillates less than the same plot of $H^u_{\textrm{val}}$, which is understandable because $H$ is built as the difference of two structures with an oscillating behavior. 

\begin{figure}[!h]
\begin{center}
\includegraphics[width = 0.8\textwidth]{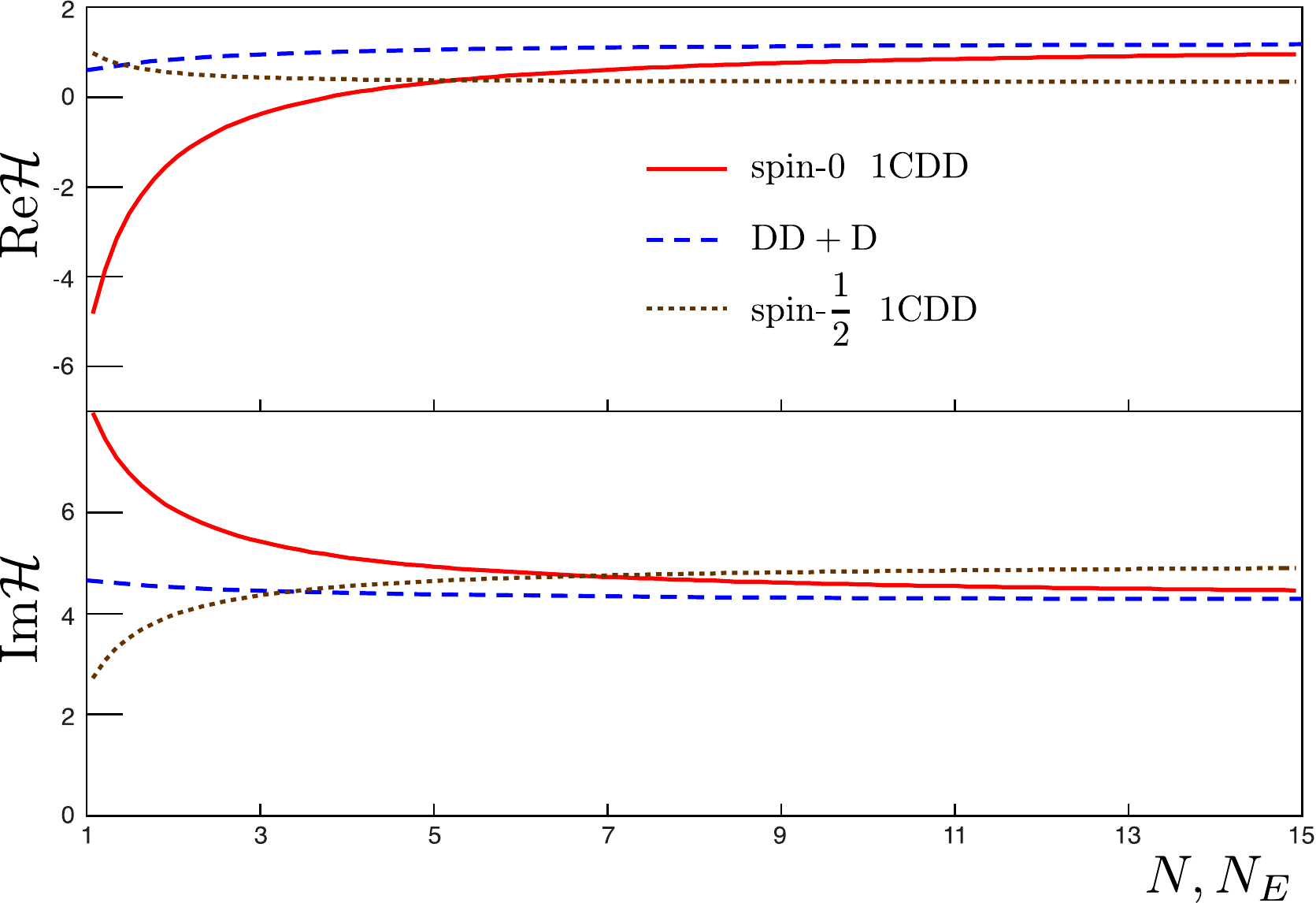}
\caption{\label{fig:CFF-nucleon-target}Comparison of flavor and charge singlet CFFs  $\mathcal{H}$ and $\mathcal{E}$ when evolving $N_E$ (spin-1/2 model discussed in this section) and $N$ (spin-0 and GK models discussed in previous sections) for $\xb = 0.36$, $t = -0.23~\GeV^2$ and $Q^2 = 2.3~\GeV^2$. The full red line corresponds to the spin-0 1CDD parameterization, the dashed blue line to the DD+D parameterization, the dotted brown line to the spin-1/2 1CDD GPD $H$.}
\end{center}
\end{figure}

\reffig{fig:CFF-nucleon-target} compares the real and imaginary parts of the charge and flavor singlet CFF $\mathcal{H}$ and $\mathcal{E}$ when varying the parameter of the profile function used in the 1CDD description for the original GK model, its modification inspired by the spinless case discussed in the first sections of this paper, and the extension of the 1CDD model to the nucleon case. The range of possible values for $\ReH$ and $\ImH$ is smaller for the spin-1/2 than for the spin-0 model. In this implementation of the nucleon model, the dependence of $\mathcal{H}$ on $N$ was probably softened by the addition of its DD part, inducing an apparent loss of flexibility. 

\subsection{Comparison to Jefferson Lab measurements}

\begin{figure}[!h]
\begin{center}
\includegraphics[width=\textwidth]{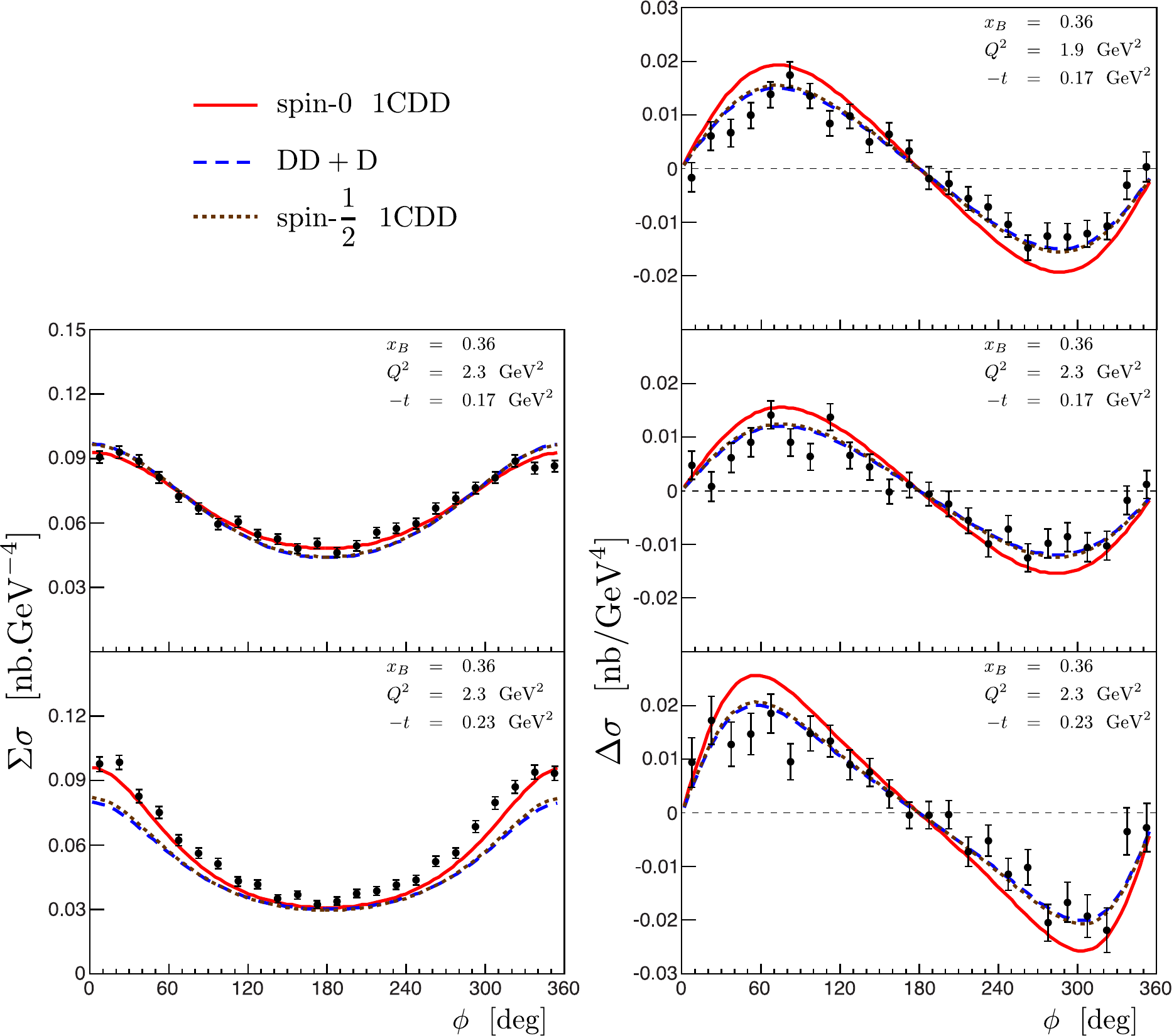}
\caption{\label{fig:results-fit-nucleon-HallA}Comparison to JLab Hall~A helicity-dependent and independent cross sections such that $\frac{|t|}{Q^2} \le 0.1$. The full red line corresponds to the spin-0 1CDD parameterization, the dashed blue line to the DD+D parameterization, and the dotted brown line to the spin-1/2 1CDD parameterization.}
\end{center}
\end{figure}

\begin{figure}[!h]
  	\begin{center}
		\begin{tabular}{c}
  		\includegraphics[width=\textwidth]{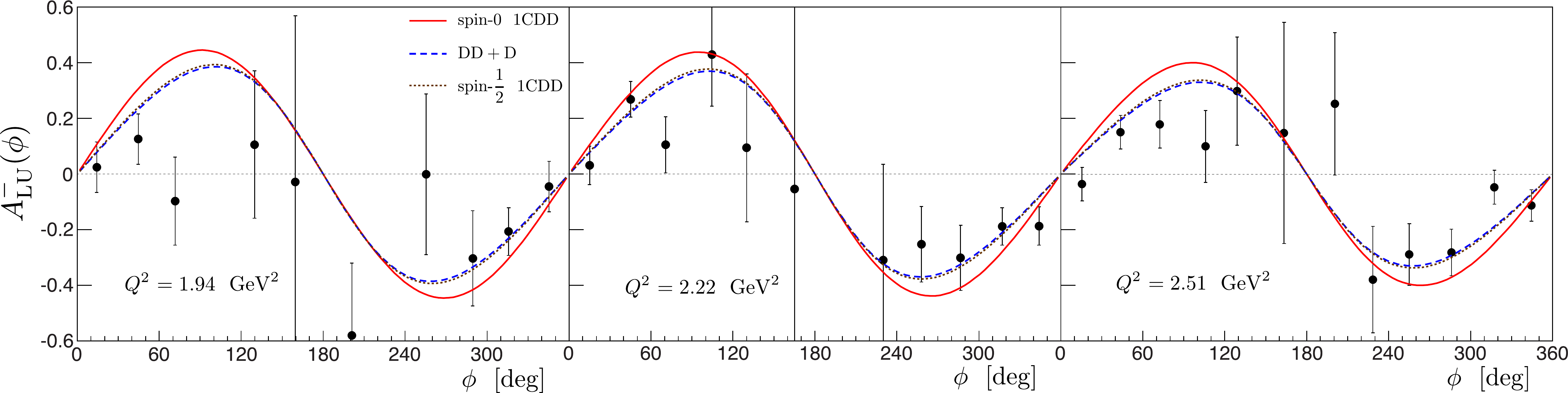} 
		\end{tabular}
  		\caption{\label{fig:Comparison-Hall-B-nucleon}Comparison of 1CDD (full red line) and DD+D (dashed blue line) models with CLAS data at $t \simeq -0.17~\GeV^2$ and such that $\frac{|t|}{Q^2} \le 0.1$. From left to right: $\xb = 0.3205$, $t = -0.1705~\GeV^2$ and $Q^2 = 1.9424~\GeV^2$; $\xb = 0.3215$, $t = -0.1719~\GeV^2$ and $Q^2 = 2.217~\GeV^2$; $\xb = 0.3215$, $t = -0.1743~\GeV^2$ and $Q^2 = 2.5078~\GeV^2$. The full red line corresponds to the spin-0 1CDD parameterization, the dashed blue line to the DD+D parameterization, and the dotted brown line to the spin-1/2 1CDD parameterization.}
  		
	\end{center}
\end{figure}

We now compare all three models discussed in this paper to Hall~A and CLAS data:
\begin{description}
\item[DD+D] GK model supplemented by a $D$-term as in \refeq{eq:D-Term-nucleon} obtained from a fit of the data.
\item[spin-0 1CDD] Adaptation of the 1CDD formalism in the spinless case to the GPD $H$ as discussed in the beginning of the paper.
\item[spin-1/2 1CDD] Combination of 1CDD and DD formalisms as detailed in this section.
\end{description} 
The three models behave differently (see \reffig{fig:results-fit-nucleon-HallA}). The spin-0 model does a better job on beam helicity-independent cross sections whereas both spin-1/2 and DD+D models offer a better agreement with beam helicity-dependent cross sections. We note that the spin-1/2 model and the DD+D model are hardly distinguishable (see \reftab{tab:chisquare-fit-nucleon-model}). This is also what we observe when comparing the three models to CLAS beam spin asymmetries (see \reffig{fig:Comparison-Hall-B-nucleon}).

This surprising result may be explained by the additional $D$-term and the lack of sensitivity of these observables to GPD $E$. In particular, the best fit we obtained corresponds to $N_{H+E} \simeq 1$ and $N_E \rightarrow \infty$. From previous CFF plots \reffig{fig-CFF-H-vs-N} and \reffig{fig:CFF-nucleon-target} we see that $N \simeq 10$ corresponds to the FPD limit with a good approximation, so $N_E$ was searched between 1 and 10, and the fit systematically gives 10. This situation will certainly improve by adding data with a higher sensitivity to $E$, but no such DVCS measurement is available yet in the valence region. HERMES provide such datasets in the intermediate $\xb$ region but taking them into account would require the parameterization of the sea in the 1CDD framework and the treatment of a PDF more divergent at small nucleon momentum fractions. 

\begin{table}[h]
\begin{center}
\begin{tabular}{|c|c|c|}
\hline
\hline
model 			& $t = -0.17~\GeV^2$ 		& $t=-0.23~\GeV^2$ \\
\hline
DD+D 			& 1.9 					& 10.0 \\
spin-0 1CDD 		& 2.6 					& 4.1 \\
spin-1/2 1CDD 	& 1.9 					& 9.4 \\
\hline
\hline
\end{tabular}
\caption{\label{tab:chisquare-fit-nucleon-model}$\chi^2$ per degree of freedom for the different models and for the different values of $t$.}
\end{center}
\end{table}

\section*{Conclusion}

We have discussed the One and Two-Component DD representations and explained why the Factorized Double Distribution Ansatz breaks their equivalence. We compared these two representations to existing data. To keep the exercise simple, we worked with beam-helicity dependent observables in the valence region. As a first approximation, we choose to work at leading order, to take into account only the quark GPDs $H$ and $E$, and to modify only their valence part. We also neglect GPD evolution and higher twist effects. Such a detailed treatment is beyond the scope of this paper, which simply aims at comparing the pros and cons of two DD representations. 

To illustrate the Two-Component DD representations (spin-0 and spin-1/2 cases) we used the Goloskokov~-~Kroll model. We built our One-Component DD models by modifying the valence parts of $H$ and $E$ and leaving the rest unchanged. Interestingly, for a given GPD, all models have the same unskewed limits when the profile function widths decrease to produce a single peak. This shows a natural limit of DD models and gives a hint about the flexibility of the associated DD representations. Since both One and Two-Component DD representations have the same limit when distorting the profile function, the One-Component DD framework can produce similar results to the Two-component DD representation.

The classical Two-Component DD is almost insensitive to the width of the profile function, while the One-Component DD displays important variations when using a spin-0 Ansatz. The conclusion is less clear when using a spin-1/2 modeling, which mixes $H$ and $E$ in the DD and One-Component DD representations, probably because the $E$-dependent part of the model is not really constrained by the selected measurement sets. In that respect it would be interesting to carefully examine the constraints on $E$ brought by the form factor $F_2$ in the spirit of the extensive study \refcite{Diehl:2013xca}. This point is left for future work. 

The One-Component DD has hardly been used so far in phenomenological applications because of its more singular behavior for small longitudinal momentum fractions. But since Radyushkin's treatment of divergences \cite{Radyushkin:2011dh} in the One-Component DD representation, the implementation of the One-Component DD and Two-Component DD frameworks have the same complexity, at last when the GPD forward limits have integrable singularities: the same kind of integrals have to be dealt with. 

For these two reasons (flexibility and implementation complexity) we consider the use of the One-Component DD representation and its implementation along the lines described in \refcite{Radyushkin:2011dh, Radyushkin:2013hca} as an interesting alternative to the earlier approach of \refcite{Radyushkin:1998es, Radyushkin:1998bz, Musatov:1999xp}. These features allow to build flexible ands realistic GPD models for sensitivity studies, for example to compute the typical size of higher-order corrections in some channels in the spirit of \refcite{Moutarde:2013qs}.

In this study GPDs were modified only in the valence region. It would be very useful to extend the One-Component DD implementation in a way to regularize the divergences associated with sea PDFs.


\section*{Acknowledgments}

The authors thank K.~Semenov-Tian-Shanskii for discussions at the initial stage of this work, A.~Radyushkin for several enlightening discussions about Generalized Parton Distributions and Double Distributions, and M.~Diehl for useful comments.

This work is partly supported by the Commissariat \'a l'Energie Atomique, the Ecole Normale Sup\'erieure de Cachan, the Joint Research Activity "Study of Strongly Interacting Matter" (acronym HadronPhysics3, Grant Agreement n.283286) under the Seventh Framework Programme of the European Community, by the GDR 3034 PH-QCD "Chromodynamique Quantique et Physique des Hadrons", and the ANR-12-MONU-0008-01 "PARTONS".


\end{document}